\documentclass[12pt]{article}
\usepackage{amssymb}
\usepackage{cite}
\newcommand{\be}{\begin{eqnarray}}
\newcommand{\ee}{\end{eqnarray}}
\newcommand{\bez}{\begin{eqnarray*}}
\newcommand{\eez}{\end{eqnarray*}}

\newcommand{\na}{\nabla}
\renewcommand{\d}{{\rm d}}
\newcommand{\dl}{\delta}
\newcommand{\ad}{{\rm ad}}
\newcommand{\A}{{\cal A}}
\newcommand{\X}{{\cal X}}

\newcommand{\oA}{\otimes_\A}
\newcommand{\oL}{\otimes_L}
\newcommand{\V}{{\cal V}}

\newcommand{\E}{\mathfrak{E}}
\newcommand{\g}{\mathfrak{g}}
\newcommand{\R}{{\cal R}}
\newcommand{\tth}{\tilde{\theta}}

\title{\bf Automorphisms of associative algebras  \\
           and noncommutative geometry}
\date{  }
\author{A. Dimakis$^1$ and F. M\"uller-Hoissen$^2$}

\begin{document}
\renewcommand{\theequation} {\arabic{section}.\arabic{equation}}
\maketitle

\newtheorem{lemma}{Lemma}[section]

\newcounter{example}[section]
\renewcommand{\theexample}{\arabic{section}.\arabic{example}}
\newenvironment{example}
{ \refstepcounter{example} \noindent
  {\it Example \arabic{section}.\arabic{example}}.}
{ \vspace{.2cm} }

\newenvironment{app_example}
{ \refstepcounter{example} \noindent
  {\it Example \Alph{section}.\arabic{example}}.}
{ \vspace{.2cm} }

\begin{center}
 $^1$ Department of Financial and Management Engineering,
 University of the Aegean, \\
 31 Fostini Street, GR-82100 Chios, dimakis@aegean.gr
\vskip.1cm
 $^2$ Max-Planck-Institut f\"ur Str\"omungsforschung,
 Bunsenstrasse 10, D-37073 G\"ottingen,
 fmuelle@gwdg.de
\end{center}

\begin{abstract}
A class of differential calculi is explored which is determined
by a set of automorphisms of the underlying associative algebra.
Several examples are presented. In particular, differential calculi
on the quantum plane, the $h$-deformed plane and the quantum
group $GL_{p,q}(2)$ are recovered in this way.
Geometric structures like metrics and compatible linear connections
are introduced.
\end{abstract}

\section{Introduction}
\setcounter{equation}{0}
In this work $\A$ denotes an associative algebra, with unit $\mathbf{1}$,
over a field $\Bbbk$ of characteristic zero (typically $\mathbb{R}$ or
$\mathbb{C}$), and $(\Omega(\A),\d)$ a differential calculus over $\A$.
In many relevant examples in mathematics and physics (see, e.g.,
Ref.~\citen{Mado99} for an introduction) the commutation relations between
elements $f \in \A$  and 1-forms can be expressed as
\be
    \theta^s \, f = \sum_{s' \in S} \Phi(f)^s_{s'} \; \theta^{s'}
                    \qquad \forall s \in S   \label{theta_Phi}
\ee
with respect to a left and right $\A$-module basis of $\Omega^1(\A)$.
Here $S$ is some finite set.
Linearity and associativity then require that $f \mapsto \Phi(f)$
is an algebra isomorphism. If the basis
$\{ \theta^s \, | \, s \in S \}$ can be chosen
in such a way that $\Phi$ is diagonal\footnote{Let us consider a change
of basis $\theta^s \mapsto {\theta'}^s := \sum_{s' \in S} U^s{}_{s'} \, \theta^{s'}$
where $U$ is an invertible matrix with entries in $\A$, or some
extension of $\A$. Then (\ref{theta_Phi}) holds with the substitution
$\Phi(f) \mapsto \Phi'(f) := U \, \Phi(f) \, U^{-1}$.
The problem is to find a $U$ such that $\Phi'(f)$ is
diagonal for all $f \in \A$.}, we obtain
\be
   \theta^s \, f = \phi_s(f) \, \theta^s    \label{theta_f_phi}
\ee
and the maps $\phi_s$ are {\em automorphisms} of $\A$.
\vskip.1cm

It is not always possible to achieve the special structure (\ref{theta_f_phi}).
Even if it is possible, then there is in general a price to pay for it:
either we have to allow for ``generalized'' differential calculi,
or we have to extend the algebra $\A$ (and the differential calculus
over it).
\vskip.1cm

A \emph{differential calculus} over $\A$ is an $\mathbb{N}_0$-graded
associative algebra $\Omega(\A)=\bigoplus_{r \geq 0} \Omega^r(\A)$ with
$\Omega^0(\A) = \A$ and $\A$-bimodules $\Omega^r(\A)$, together with
a $\Bbbk$-linear map $\d : \Omega^r(\A) \rightarrow \Omega^{r+1}(\A)$
satisfying $\d^2 = 0$ and the Leibniz rule
\be
 \d (\omega \, \omega') = (\d \omega) \, \omega'
                       + (-1)^r \, \omega \, \d \omega'
                       \label{d_Leibniz}
\ee
for all $\omega \in \Omega^r(\A)$ and $\omega' \in \Omega(\A)$. We require
that $\mathbf{1}$ is also a unit of $\Omega(\A)$ (which implies
$\d \mathbf{1} = 0$). In addition one usually demands that
$\Omega^{r+1}(\A)$ coincides with the $\A$-bimodule generated by $\d \Omega^r(\A)$.
We shall admit, however, the possibility that the space of 1-forms is
larger than the $\A$-bimodule generated by $\d \A$. In this case
we talk about a \emph{generalized differential calculus} and denote it
as $(\hat{\Omega}(\A),\d)$. Of course, it always contains a sub-differential
calculus $(\Omega(\A),\d)$ which is an ordinary differential calculus.
\vskip.1cm

There are in fact prominent examples for which the structure
(\ref{theta_f_phi}) appears.
Consider a commutative algebra $\A$ with an automorphism $\phi$ and
the universal differential calculus $(\Omega_u(\A),\d_u)$ over $\A$.
Let $(\Omega_\phi(\A), \d_\phi)$ be the quotient of the universal
differential graded algebra $\Omega_u(\A)$ by the ideal generated
by the relations $(\d_u a) \, b - \phi(b) \, \d_u a$ for all
$a,b \in \A$. If $\phi$ is the identity, one recovers the K\"ahler
differentials \cite{Kaehler}.
Otherwise one obtains ``twisted K\"ahler differentials'' \cite{Karoubi}.
A particular example \cite{qcalc,Karoubi} is given by
$\A = \Bbbk[x]$ with $\phi(f(x)) = f(q x)$ for all $f \in \A$ and
$q \in \Bbbk\setminus \{ 0 \}$, so that $\d x \, f = \phi(f) \, \d x$.
The structure expressed in (\ref{theta_f_phi}) generalizes this
construction in several ways: more than a single automorphism is admitted,
the 1-forms $\theta^s$ are not required to be exact, the algebra $\A$
need not be commutative.
\vskip.1cm

Another class of examples is given by group lattice differential
calculi \cite{DMH02a,DMH02b} which are determined by a discrete group
$G$ and a finite subset $S$ not containing the unit element.
The 1-forms $\theta^s$, $s \in S$, are dual to the
discrete derivatives
\be
    e_s = \phi_s - \mbox{id}  \label{vf_e_s}
\ee
acting on the algebra of functions on $G$. Here $\phi_s = R_s^\ast$
is the pull-back with the right action $R_s(g)=gs$ (for all $g \in G$).
This structure generalizes as follows. Given a set of automorphisms
$\phi_s$ of an associative algebra $\A$, (\ref{vf_e_s}) defines
``vector fields'' with which we associate, via duality relations,
a set of objects $\theta^s$ so that (\ref{theta_f_phi}) holds.
It turns out that a (generalized) differential calculus
can indeed be constructed in this way.
An interesting aspect is that such differential calculi
are naturally associated with the underlying algebra $\A$, since
they only use properties of $\A$ and no ``external'' structures.
More precisely, for this construction it is required that the
automorphisms $\phi_s$ are sufficiently different from each other
and different from the identity. On the other hand,
there are examples of differential calculi over noncommutative
algebras where (\ref{theta_f_phi}) holds with $\phi_s = \mbox{id}$
for some or even all $s \in S$, see Ref.~\citen{Dima+Mado96}.
Such calculi may be recovered as limits of curves of the kind
of calculi described above (see example~\ref{ex:matrix}).
But this also suggests a further generalization of the discrete
derivatives (\ref{vf_e_s}) to ``twisted inner derivations", see
(\ref{twist_deriv}).
\vskip.1cm

It turns out that the (generalized) differential calculi
obtained in this way are {\em inner} at first order.
Section~\ref{sec:dc_inner} provides some information about
calculi possessing this property.
\vskip.1cm

Section~\ref{sec:diff_auto} deals with homomorphisms
$\A \rightarrow \A$ which are ``differentiable'' with respect
to a differential calculus.
Section~\ref{sec:dc_auto} introduces a class of differential calculi
associated with automorphisms of the underlying algebra $\A$ in the
sense sketched above, based on the discrete derivatives (\ref{vf_e_s}).
Several examples are presented.
The generalization based on twisted inner derivations is then
the subject of section~\ref{sec:general} where in particular
an example of a bicovariant differential calculus \cite{Woro89,Klim+Schm97}
on the quantum group $GL_{p,q}(2)$ \cite{MH92} is treated.
\vskip.1cm

In section~\ref{sec:conn} we recall some facts about connections
and, in particular, linear connections.
Section~\ref{sec:ncg} treats differential calculi associated
with differentiable automorphisms in which case a (semi-) left-linear
tensor product and thus close analogues of the tensors of classical
differential geometry can be defined.
Finally, section~\ref{sec:conclusions} contains some concluding remarks.

\section{Differential calculi which are inner at first order}
\label{sec:dc_inner}
\setcounter{equation}{0}
Let $(\hat{\Omega}(\A),\d)$ be a (possibly generalized) differential calculus
over $\A$ such that
\be
    \d f = [\vartheta,f]  \qquad \forall f \in \A  \label{d_inner}
\ee
with a 1-form $\vartheta \in \hat{\Omega}^1(\A)$. In this case the first
order differential calculus $\d : \A \rightarrow \hat{\Omega}^1(\A)$ is said
to be ``inner''. Of course, we may add to $\vartheta$ any 1-form which
commutes with all elements of $\A$.
Inner differential calculi appear frequently in the literature. In particular,
it seems that most bicovariant differential calculi on quantum groups
have this property, see Refs.~\citen{dc_qg,Klim+Schm97}.\footnote{As
demonstrated in Ref.~\citen{Woro89}, the bimodule of a bicovariant first
order differential calculus can always be extended with an additional
generator so that the extended bimodule is also bicovariant and the
differential calculus becomes inner. We then have a generalized differential
calculus in the sense of the introduction.}
There are also several examples of inner first order differential calculi on
quantum homogeneous spaces \cite{Schm98}.
\vskip.1cm

The expression
\be
   \Delta(\omega) := [\vartheta,\omega] - \d \omega  \label{Delta_def}
\ee
involving the graded commutator defines a linear map
$\Delta : \hat{\Omega}(\A) \rightarrow \hat{\Omega}(\A)$. As a consequence
of the Leibniz rule (\ref{d_Leibniz}) and the properties of the
commutator, it satisfies
\be
    \Delta(\omega \omega')
  = \Delta(\omega) \, \omega' + (-1)^r \omega \, \Delta(\omega')
    \label{Delta_deriv}
\ee
for all $\omega \in \hat{\Omega}^r(\A)$ and $\omega' \in \hat{\Omega}(\A)$.
Hence $\Delta$ is a graded derivation of $\hat{\Omega}(\A)$ of grade 1.
Moreover, (\ref{d_inner}) implies $\Delta(f) = 0$ for all $f \in \A$,
so that $\Delta$ is an $\A$-bimodule homomorphism, i.e.
\be
     \Delta (f \omega f') = f \Delta(\omega) f'
     \qquad \forall f,f' \in \A, \, \omega \in \hat{\Omega}(\A) \; .
\ee
\vskip.1cm

The 2-form
\be
    \zeta := \d \vartheta - \vartheta^2 = \vartheta^2 - \Delta(\vartheta) \label{zeta}
\ee
commutes with all elements of $\A$:
\be
      [\zeta,f]
  &=& [\d \vartheta , f] - [\vartheta^2,f]
   = \d [\vartheta,f] + [\vartheta,\d f] - [\vartheta,[\vartheta,f]] \nonumber \\
  &=& \d \d f + [\vartheta,\d f] - [\vartheta,\d f] = 0 \; . \label{zeta_f}
\ee
Furthermore, acting with $\Delta$ on (\ref{Delta_def}), using
(\ref{Delta_deriv}) and $\d^2 =0$, we find
\be
   \Delta^2 (\omega) = [ \Delta (\vartheta), \omega ] - [\vartheta, [\vartheta,\omega]]
   = [ \Delta (\vartheta), \omega ] - [\vartheta^2, \omega]
   = - [ \zeta, \omega ] \; .  \label{Delta^2}
\ee
An immediate consequence is the following.

\begin{lemma}
$\Delta^2 = 0$ if and only if $\zeta$ lies in the center of $\hat{\Omega}(\A)$.
\hfill $\blacksquare$
\end{lemma}

Using (\ref{Delta^2}), we obtain
\be
   \Delta(\zeta) = \Delta(\vartheta^2) - \Delta^2(\vartheta)
   = [\Delta(\vartheta) + \zeta, \vartheta] = [\vartheta^2,\vartheta] = 0 \, ,
\ee
so that (\ref{Delta_def}) implies
\be
    \d \zeta = [ \vartheta , \zeta ] \; .
\ee
\vskip.1cm

A distinguished case occurs if $\Delta$ vanishes on $\hat{\Omega}(\A)$. Then the
differential calculus is also inner at higher orders.

\section{Differentiable maps}
\label{sec:diff_auto}
\setcounter{equation}{0}
A homomorphism $\phi$ of $\A$ is called \emph{differentiable} with respect
to a (possibly generalized) differential calculus $(\hat{\Omega}(\A),\d)$
if it extends to a homomorphism of $\hat{\Omega}(\A)$ (as a graded algebra
over $\Bbbk$) such that
\be
    \phi \circ \d = \d \circ \phi \; .
\ee
\vskip.1cm

Let $(\Omega(\A),\d)$ be the maximal ordinary differential calculus contained
in $(\hat{\Omega}(\A),\d)$.
Since, by use of the Leibniz rule, every element $\omega \in \Omega^r(\A)$
can be expressed as
\be
    \omega = \sum_{a=1}^N f_{(0)a} \, \d f_{(1)a} \ldots \d f_{(r)a} \, ,
    \label{omega_df}
\ee
with $f_{(p)a} \in \A$, we may think of extending a homomorphism
$\phi$ of $\A$ simply by defining
\be
   \phi(\omega) := \sum_{a=1}^N \phi(f_{(0)a}) \, \d \phi(f_{(1)a})
                   \ldots \d \phi(f_{(r)a})  \; .
\ee
At least, this is consistent if the differential calculus is the universal
one. If not, then there are special linear combinations of forms
which vanish identically. But the right hand side of the above definition
need not respect this and would then lead to a contradiction.
\vskip.1cm

In this work, we restrict our considerations to differential calculi
which are generated by first-order differential calculi without imposing
further relations by hand\footnote{An example for such a further relation
would be the Woronowicz wedge product \cite{Woro89} which is stronger,
in general, than what is required to extend a (bicovariant) first order
differential calculus to higher orders.}
on the level of $r$-forms with $r>1$.
The differentiability condition for a homomorphism $\phi$ of $\A$
with respect to $(\Omega(\A),\d)$ then becomes
\be
   \sum_{a=1}^N f_a \, \d f_a' = 0 \quad \Longrightarrow \quad
   \sum_{a=1}^N \phi(f_a) \, \d \phi(f_a') = 0
                              \label{phi_diff_criterion}
\ee
for all $f_a,f_a' \in \A$ and all $N \in \mathbb{N}$. In the case of
a generalized differential calculus $(\hat{\Omega}(\A),\d)$, there are further
relations which $\phi$ has to preserve in order to be differentiable.
\vskip.1cm

Most examples of generalized differential calculi considered in this
work (see section~\ref{sec:dc_auto}) are actually of a rather special kind.
They are obtained by \emph{minimal extensions} of $\Omega^1(\A)$ in the following way.
We say that an element $\alpha \in \hat{\Omega}^1(\A) \setminus \Omega^1(\A)$
\emph{minimally extends} $\Omega^1(\A)$ if there
is an $f \in \A \setminus \{ 0 \}$ such that $f \, \alpha \in \Omega^1(\A)$
and $f$ has no inverse in $\A$. Then $f \, \alpha = \sum_a f_a \, \d f_a'$
with $f_a, f_a' \in \A$.
In order to extend $\phi$ to $(\hat{\Omega}(\A),\d)$ and thus to achieve that
$\phi$ is differentiable with respect to the generalized differential
calculus, we have to define $\phi(\alpha)$ in accordance
with $\phi(f) \, \phi(\alpha) = \sum_a \phi(f_a) \, \d \phi(f_a')$.
\vskip.1cm

A first order differential calculus $\d : \A \rightarrow \hat{\Omega}^1(\A)$ is
called \emph{simple} if there is no (non-vanishing) 1-form which commutes
with all elements of $\A$.

\begin{lemma}
\label{lemma:phi}
Let $\d : \A \rightarrow \hat{\Omega}^1(\A)$ be inner and simple, and $\phi$
a differentiable automorphism of $\A$. Then
\be
    \phi (\vartheta) &=& \vartheta \, ,  \\
    \phi \circ \Delta &=& \Delta \circ \phi  \; .
\ee
\end{lemma}
{\bf Proof:} Using (\ref{d_inner}) we obtain
$[ \phi (\vartheta) , \phi(f) ] = \phi( \d f ) = \d \phi(f)
 = [ \vartheta , \phi(f) ]$ for all $f \in \A$ which implies that $\phi (\vartheta) - \vartheta$
is a 1-form which commutes with all $f \in \A$. But we assumed that the
first order differential calculus is simple so that such a 1-form does not exist.
Hence $\phi (\vartheta) - \vartheta = 0$. Furthermore, applying $\phi$ to (\ref{Delta_def})
leads to $\phi(\Delta(\omega)) = [ \phi(\vartheta), \phi(\omega)] - \d \phi(\omega)
= [ \vartheta, \phi(\omega) ] - \d \phi(\omega) = \Delta( \phi(\omega) )$ for
all $\omega \in \hat{\Omega}(\A)$.
\hfill $\blacksquare$

\vskip.2cm
\noindent
{\em Remark.} If the first order differential calculus is not simple,
the Lemma can be generalized as follows. Let ${\cal Z}(\hat{\Omega}^1(\A),\A)$
denote the set of 1-forms which commute with all elements of $\A$. Then
\be
  \phi(\vartheta) = \vartheta + \vartheta_\phi \, , \quad
  \phi \circ \Delta(\omega) = \Delta \circ \phi (\omega) + [\vartheta_\phi,\phi(\omega)] \, ,
\ee
with some $\vartheta_\phi \in {\cal Z}(\hat{\Omega}^1(\A),\A)$, depending on $\phi$.
\hfill $\blacksquare$

\section{A class of differential calculi determined by automorphisms}
\label{sec:dc_auto}
\setcounter{equation}{0}
Let $G$ be a group and $S \subset G$ a finite subset which acts faithfully
by automorphisms $\{ \phi_s \, | \, s \in S \}$ on an associative algebra $\A$,
so that $\phi_s(f) = \phi_{s'}(f)$ for all $f \in \A$ implies $s=s'$.
We define maps $e_s : \A \rightarrow \A$ such that
\be
    e_s f = [ \phi_s(f) - f ]/t_s  \qquad (s \in S)  \label{e_s}
\ee
with an arbitrary parameter $t_s \in \Bbbk$.\footnote{If $\phi_s$ depends on the
parameter $t_s$ and if a norm is defined on $\A$, it may be of interest to
consider the limit $t_s \to 0$. Otherwise we may set $t_s=1$ without restriction
of generality.}
As a consequence, the operators $e_s$ satisfy a twisted derivation rule:
\be
   e_s(f f') = (e_s f) \, \phi_s(f') + f \, (e_s f') \; .
   \label{twder}
\ee
The linear span of mappings $\{ e_s \cdot f \, | \, s \in S, \, f \in \A \}$,
where
\be
    (e_s \cdot f) \, f' = (e_s f') \, f \, ,  \label{ef1}
\ee
is turned into an $\A$-bimodule $\X$ via
\be
  f \cdot e_s = e_s \cdot \phi_s(f) \; .  \label{ef2}
\ee
The elements of $\X$ will be called \emph{vector fields}.
In the following, we assume that $\X$ is free with $\{ e_s | s \in S \}$
as a left and right $\A$-module basis.
\vskip.1cm
\noindent
{\em Remark.} Let us consider the algebra $\A$ obtained from $\mathbb{C}[x]$,
the algebra of polynomials in a variable $x$ with coefficients in $\mathbb{C}$,
by imposing the relation $x^2 = 0$.
Then $\phi_s(x) = \mu_s \, x$, $s=1,2$, where $\mu_s \in \mathbb{C} \setminus \{ 0,1 \}$
are such that $\mu_1^k \neq \mu_2^l$ for all $k,l \in \mathbb{Z} \setminus \{ 0 \}$,
determines a faithful action of $\mathbb{Z}^2$ by automorphisms.
$e_1 \cdot f_1 + e_2 \cdot f_2 = 0$ is satisfied with $f_2 = -f_1 (\mu_1-1)/(\mu_2-1)$
and arbitrary $f_1 \in \mathbb{C}$. Hence $\{ e_s | s=1,2 \}$ is not a right $\A$-module
basis. This example shows that a faithful action does not, in general, guarantee
that $\{ e_s \}$ is a left and right $\A$-module basis. In many interesting examples,
the relations in $\A$ are much less restrictive, however, and the latter
property usually holds.
\hfill $\blacksquare$
\vskip.1cm

Let $\hat{\Omega}^1(\A)$ be the dual $\A$-bimodule obtained with a contraction which
has the following properties:
\be
  \langle \alpha f , X \rangle &=& \langle \alpha , f \cdot X \rangle \label{dual1} \\
  \langle \alpha , X \cdot f \rangle &=& \langle \alpha , X \rangle \, f  \\
  \langle f \alpha , X \rangle &=& f \, \langle \alpha , X \rangle \label{dual3}
\ee
for all $f \in \A$, $X \in \X$ and $\alpha \in \hat{\Omega}^1(\A)$. Now
\be
   \langle \theta^s , e_{s'} \rangle = \delta^s_{s'}
\ee
defines duals of $e_s$, $s \in S$. Using (\ref{dual1}) and (\ref{ef2}) we obtain
\be
     \langle \theta^s f , e_{s'} \rangle
 &=& \langle \theta^s , f \cdot e_{s'} \rangle
  = \langle \theta^s , e_{s'} \cdot \phi_{s'}(f) \rangle
  = \langle \theta^s , e_{s'} \rangle \, \phi_{s'}(f)
  = \delta^s_{s'} \, \phi_{s'}(f)    \nonumber \\
 &=& \langle \phi_s(f) \, \theta^s , e_{s'} \rangle
\ee
so that
\be
   \theta^s \, f = \phi_s(f) \, \theta^s \; .
   \label{cmr_thetaf}
\ee
Next we define a linear map $\d : \A \rightarrow \hat{\Omega}^1(\A)$ by
\be
     X f = \langle \d f , X \rangle
\ee
for all $X \in \X$ and $f \in \A$. This implies
\be
    \d f = \sum_{s \in S} (e_s f) \, \theta^s \; .  \label{d_e}
\ee
As a consequence of (\ref{twder}) and (\ref{cmr_thetaf}), $\d$
satisfies the Leibniz rule
\be
   \d (f f') = (\d f) \, f'+ f \, (\d f')
\ee
so that $(\hat{\Omega}^1(\A),\d)$ constitutes a (possibly generalized)
first order differential calculus.
\vskip.1cm

In general, it is \emph{not} possible to express $\theta^s$ as
a linear combination of differentials of elements of $\A$
with coefficients in $\A$, as in (\ref{omega_df}), i.e.,
$\theta^s$ need not be an element of $\Omega^1(\A)$.
\vskip.1cm

If there are elements $f^s \in \A$, $s \in S$, such that
$e_s f^{s'}$ is an invertible matrix and the inverse has entries
in $\A$, then we can solve (\ref{d_e}) for the $\theta^s$ and they
lie in $\Omega^1(\A)$ (see example~\ref{example:Zs}).
In this case, the elements $f^s$ of $\A$ may be regarded
as \emph{coordinates}.

\begin{lemma}
\label{lemma:simple}
A 1-form $\alpha = \sum_{s \in S} \alpha_s \, \theta^s$ with
$\alpha_s \in \A$ commutes with all elements of $\A$ iff
\be
   \alpha_s \, \phi_s(f) = f \, \alpha_s
   \qquad \forall f \in \A , \quad \forall s \in S \; .
\ee
Furthermore, if $\phi_s(f) = U_s \, f \, U_s^{-1}$ with an invertible
$U_s \in \A$, so that $\phi_s$ is an inner automorphism,
then $\alpha_s U_s$ belongs to the center of $\A$.
\end{lemma}
{\bf Proof:} The first statement is a direct consequence of (\ref{cmr_thetaf}).
The second immediately follows from the first, using the special form of $\phi_s$.
\hfill $\blacksquare$
\vskip.2cm

The Lemma gives a criterion for the first order differential calculus
$\d : \A \rightarrow \hat{\Omega}^1(\A)$ to be simple.
\vskip.1cm

Using (\ref{e_s}), (\ref{cmr_thetaf}) and (\ref{d_e}), (\ref{d_inner})
holds with
\be
  \vartheta = \sum_{s \in S} {1 \over t_s} \, \theta^s  \label{vartheta_t}
\ee
so that all formulas of section~\ref{sec:dc_inner} apply in
the case under consideration.
\vskip.2cm

Differentiability of a homomorphism $\phi : \A \rightarrow \A$ with
respect to $(\Omega(\A), \d)$ means that the criterion (\ref{phi_diff_criterion})
has to be fulfilled.
Differentiability of $\phi$ with respect to a generalized differential
calculus $(\hat{\Omega}(\A), \d)$ requires in addition that it must be
possible to define $\phi$ on $\theta^s$ consistently in such a way
that (\ref{d_e}) is preserved, i.e.,
$\d \phi(f) = \sum_{s \in S} \phi(e_s f) \, \phi(\theta^s)$
for all $f \in \A$. Furthermore, (\ref{cmr_thetaf}) has to be preserved
which leads to
\be
  \phi(\theta^s) \, f = \phi \circ \phi_s \circ \phi^{-1}(f) \, \phi(\theta^s) \; .
\ee
\vskip.1cm

If $\phi_s$ is differentiable with respect to $(\hat{\Omega}(\A), \d)$, then
\be
   \phi_s(\theta^{s'}) \, f = \phi_{ss's^{-1}}(f) \, \phi_s(\theta^{s'})
\ee
for all $s' \in S$ and all $f \in \A$. If $\mbox{ad}(s) S \subset S$,
then $\phi_s(\theta^{s'})$ satisfies the same commutation relations as
$\theta^{ss's^{-1}}$ with elements of $\A$.
We may then expect generically that $\phi_s(\theta^{s'})$ coincides with
$\theta^{ss's^{-1}}$ up to a factor which lies in ${\cal Z}(\A)$,
the center of $\A$. A corresponding criterion is given next.

\begin{lemma}
A 1-form $\alpha = \sum_{s' \in S} \alpha_{s'} \, \theta^{s'}$ satisfies
$\alpha \, f = \phi_s(f) \, \alpha$ for some fixed $s \in S$ and all $f \in \A$
iff  $\alpha_s \in {\cal Z}(\A)$ and
$\alpha_{s'} \, \phi_{s'}(f) = \phi_s(f) \, \alpha_{s'}$
for all $s' \in S \setminus \{ s \}$ and all $f \in \A$.
\end{lemma}
{\bf Proof:} This is an immediate consequence of (\ref{cmr_thetaf}).
\hfill $\blacksquare$
\vskip.1cm

An element $c \in \A$ with the property $\phi_s(c) = c$ for all $s \in S$ is
a \emph{constant} of the differential calculus: $\d c = 0$.
The ideal ${\cal J}_c$ of $\hat{\Omega}(\A)$, generated by $c - \lambda \, \mathbf{1}$
with $\lambda \in \Bbbk$, is a differential ideal, and $(\hat{\Omega}(\A)/{\cal J}_c,\d)$
is again a differential calculus, but with less constants than the one we started with.
\vskip.2cm

Every associative algebra admits a maximal first order differential
calculus, the universal first order differential calculus
$\d_u : \A \rightarrow \Omega^1_u(\A)$.
For each first order differential calculus $\d : \A \rightarrow \Omega^1(\A)$
over $\A$ there is an $\A$-bimodule homomorphism
$\pi_1 : \Omega^1_u(\A) \rightarrow \Omega^1(\A)$ such that $\d = \pi_1 \circ \d_u$.
The universal first order differential calculus extends to
the universal differential calculus $(\Omega_u(\A),\d_u)$.
Let ${\cal J}$ be the differential ideal of $\Omega_u(\A)$ generated by
$\mathrm{ker} \, \pi_1$ and let
\be
    \pi : \Omega_u(\A) \rightarrow \Omega(\A) := \Omega_u(\A)/{\cal J}
\ee
be the corresponding projection. We define $\d$ on $\Omega(\A)$ such
that $\d \circ \pi = \pi \circ \d_u$. Since ${\cal J}$ is graded,
$\pi$ is a graded homomorphism of the corresponding graded algebras.
Then $(\Omega(\A),\d)$ extends the first order differential calculus
$\d : \A \rightarrow \Omega^1(\A)$ to higher orders.
\vskip.1cm

A generalized differential calculus is \emph{not} a quotient of the
universal differential calculus. In this case a corresponding
extension is somewhat more subtle. But such problems are in some cases
easily circumvented by appending $\A$ (perhaps after abstraction from
a concrete realization) with suitable inverses of some of its elements
such that the generalized differential calculus becomes an ordinary one
over the extended algebra $\hat{\A}$.
If a resulting formula, though obtained via manipulations with inverses,
does not explicitly refer to inverses, it can be consistently
restricted to the original algebra $\A$. There are certainly other
examples which are not so easily treated.

\subsection{The structure of 2-forms and beyond}
Let us introduce the 2-forms
\be
   \zeta^g := \sum_{s,s' \in S} \dl^g_{s s'} \, {1 \over t_s t_{s'}} \, \theta^s \theta^{s'}
   \qquad (g \in G) \, .
\ee
As a consequence of (\ref{cmr_thetaf}), they satisfy
\be
    \zeta^g \, f = \phi_g(f) \, \zeta^g  \qquad \forall f \in \A \; .
\ee
In particular, $\zeta^e$ commutes with all $f \in \A$.
\vskip.1cm

In order to further analyze the structure of 2-forms, we take a closer look at
equation (\ref{zeta}). In expanded form, using (\ref{vartheta_t}), it becomes
\be
    \sum_{s,s' \in S} {1 \over t_s t_{s'}} \, \theta^s \theta^{s'}
  = \sum_{s \in S} {1 \over t_s} \, \Delta(\theta^s) + \zeta \; .
\ee
Multiplying from the right with $f \in \A$ and using the commutation relations
(\ref{cmr_thetaf}) leads to
\be
  \sum_{s,s' \in S} {1 \over t_s t_{s'}} \, \phi_{ss'}(f) \, \theta^s \theta^{s'}
 = \sum_{s \in S} {1 \over t_s} \, \phi_s(f) \, \Delta(\theta^s) + f \, \zeta \; .
             \label{2form_condition}
\ee
Since this equation holds for all $f \in \A$, and since we assumed that
$\{ \phi_s \, | \, s \in S \}$ acts faithfully on $\A$, we should expect
that it implies
\be
  \zeta &=& \sum_{s,s' \in S} {1 \over t_s t_{s'}} \, \dl^e_{s s'} \, \theta^s \theta^{s'}
         = \zeta^e  \, ,                             \label{2form_zeta} \\
  \Delta(\theta^s) &=& \sum_{s',s'' \in S} {t_s \over t_{s'} t_{s''}} \, \dl^s_{s's''}
  \, \theta^{s'} \theta^{s''} \, ,                   \label{2form_Delta}
\ee
where $e$ denotes the unit element of the group, and the 2-form relations
\be
  \zeta^g = \sum_{s,s' \in S} \dl^g_{s s'} \, {1 \over t_s t_{s'}} \, \theta^s \theta^{s'}
          = 0
  \qquad \forall g \in G \setminus (S \cup \{ e \}) \; .  \label{2formrel}
\ee
Even if these relations cannot be deduced from (\ref{2form_condition}),
it is completely consistent to impose them and we will do so.
A particular consequence is that the map $\Delta$ is now fixed and thus also
the action of $\d$ on $\hat{\Omega}(\A)$ according to (\ref{Delta_def}):
\be
  \d \omega = [\vartheta,\omega] - \Delta(\omega) \; .    \label{d_omega}
\ee
\vskip.1cm

The step from (\ref{2form_condition}) to (\ref{2form_zeta})-(\ref{2formrel}) is a
decomposition of a (general) 2-form into parts corresponding
to biangles, triangles and ``quadrangles''\footnote{This nomenclature is taken from
Ref.~\citen{DMH02a}. More precisely, here the notion ``quadrangle'' comprises
all the structures different from biangles and triangles.
If $\ad(S)S \subset S$, then these are indeed quadrangles in the sense of
Ref.~\citen{DMH02a}.}:
\be
  (s,s') \in S \times S \quad \mbox{is a} \quad
  \left\{ \begin{array}{l@{\quad}c@{\quad}l}
          \mbox{biangle} & & ss'=e \\
          \mbox{triangle} & \mbox{if} & ss'=s'' \in S \\
          \mbox{``quadrangle''} & & ss'=g \in G \setminus (S \cup \{ e \})
          \end{array} \right.
\ee
Clearly, similar decompositions can be achieved for $r$-forms with $r>2$.
For a biangle, the 2-form $\theta^s \theta^{s^{-1}}$ commutes
with all $f \in \A$.

\subsection{A collection of examples}
\label{sec:examples}

\begin{example}
\label{example:Zs}
The group $\mathbb{Z}$, or even the larger additive group of complex numbers,
acts faithfully on $\A = \mathbb{C}[x]$ via automorphisms
$(\phi_i f)(x) = f(x+i)$, $i \in \mathbb{C}$.
Let $S = \{ i_1, i_2, \ldots, i_n \} \subset \mathbb{C}$.
Then $\{ e_s = \phi_s - \mbox{id} | s \in S \}$ is a left and right
$\A$-module basis of the space $\X$. We obtain
\be
    \d x^r = [(x+i_1)^r - x^r] \, \theta^{i_1} + \ldots
             + [(x+i_n)^r - x^r] \, \theta^{i_n}
             \label{Z_n_dxr}
\ee
for $r \in \mathbb{N}$. Commuting $x$ from right to left through the
expression on the right hand side, using (\ref{cmr_thetaf}), is consistent
with the left hand side because of the Leibniz rule formula
$(\d x^r) \, x = \d x^{r+1} - x^r \, \d x$.
Next we show that the $\theta^s$ can be expressed in terms of differentials.
(\ref{Z_n_dxr}) implies
\be
   \left( \begin{array}{l} \d x \\ \d x^2 \\ \vdots \\ \d x^n \end{array} \right)
 = \left( \begin{array}{cccc} i_1 & i_2 & \ldots & i_n \\
          (x+i_1)^2-x^2 & (x+i_2)^2-x^2 & \ldots & (x+i_n)^2-x^2 \\
          \vdots & \vdots &  & \vdots \\
          (x+i_1)^n-x^n & (x+i_2)^n-x^n & \ldots & (x+i_n)^n-x^n
          \end{array} \right)
   \left( \begin{array}{l} \theta^{i_1} \\ \theta^{i_2} \\ \vdots \\ \theta^{i_n}
   \end{array} \right) \; .
\ee
Let $M_S$ denote the matrix which appears on the right hand side.
Subtracting $x$ times the $k$th row from the $(k+1)$th row of $M_S$, we obtain
$\det M_S = i_1 i_2 \cdots i_n \, {\cal D}[i_1,\ldots,i_n]$ with the determinant
\bez
 && {\cal D}[i_1,\ldots,i_n] =
 \left| \begin{array}{cccc}
 1 & 1 & \cdots & 1 \\
 x+i_1 & x+i_2 & \cdots & x+i_n \\
 (x+i_1)^2 & (x+i_2)^2 & \cdots & (x+i_n)^2 \\
 \vdots & \vdots & \ddots & \vdots \\
 (x+i_1)^{n-1} & (x+i_2)^{n-1} & \cdots & (x+i_n)^{n-1}
 \end{array} \right|  \nonumber \\
 &=& \left| \begin{array}{cccc}
 i_2-i_1 & i_3-i_1 & \cdots & i_n-i_1 \\
 (x+i_2)^2-(x+i_1)^2 & (x+i_3)^2-(x+i_1)^2 & \cdots & (x+i_n)^2-(x+i_1)^2 \\
 \vdots & \vdots & \ddots & \vdots \\
 (x+i_2)^{n-1}-(x+i_1)^{n-1} & (x+i_3)^{n-1}-(x+i_1)^{n-1} & \cdots & (x+i_n)^{n-1}-(x+i_1)^{n-1}
 \end{array} \right|
\eez
This is the determinant of a matrix with the same structure as $M_S$. We can thus
proceed as above and subtract $(x+i_1)$ times the $k$th row from the $(k+1)$th row.
This leads to
\be
    {\cal D}[i_1,\ldots,i_n]
  = (i_2-i_1)(i_3-i_1) \cdots (i_n-i_1) \, {\cal D}[i_2,\ldots,i_n] \, ,
\ee
and thus
\be
   \det M_S = i_1 \cdots i_n \,(i_2-i_1) \cdots (i_n-i_1) \, (i_3-i_2)\cdots
   (i_n-i_2)\,\cdots \,(i_n-i_{n-1}) \; .
\ee
In particular, $\det M_S$ does not vanish\footnote{The $i_k$ are assumed to be mutually
different, of course.}
and does not depend on $x$. $M_S$ is therefore invertible
and the 1-forms $\theta^s$ can be expressed as linear combinations of the
differentials $\d x^1, \ldots, \d x^n$, with coefficients in $\A$.
The automorphisms $\phi_s$ are differentiable with $\phi_s(\theta^{s'}) = \theta^{s'}$.
Indeed, using the binomial formula, we see that the expressions (\ref{Z_n_dxr})
are preserved by $\phi_s$.

 For $S = \{ 1 \}$ we have a special case of Karoubi's construction \cite{Karoubi}
(see also Ref.~\citen{DMHS93}). For $S = \{ 1, 2 \}$ we obtain
\be
   \theta^1 = 2(1+x) \, \d x - \d x^2 \, , \quad
   \theta^2 = -(\frac{1}{2}+x) \, \d x + \frac{1}{2} \, \d x^2 \; .
\ee
Then (\ref{2form_Delta}) leads to $\Delta(\theta^1) = 0$ and
$\Delta(\theta^2) = (\theta^1)^2$. From (\ref{2formrel}) we deduce
$\theta^2 \theta^1 = - \theta^1 \theta^2$ and $(\theta^2)^2 = 0$.
Since there is no biangle, according to (\ref{2form_zeta}) the 2-form $\zeta$ vanishes.

Choosing $S = \{ -1, 1 \}$ we obtain
\be
   \theta^{-1} = - \frac{1}{2} \, [ (x+1)^2 - x^2 ] \, \d x
                 + \frac{1}{2} \, \d x^2 \, , \quad
   \theta^1 = \frac{1}{2} \, [ (x-1)^2 - x^2 ] \, \d x
                 + \frac{1}{2} \, \d x^2 \, ,
\ee
so that $\vartheta = \d x^2 - 2 x \, \d x = [\d x , x]$.
Now (\ref{2formrel}) implies $(\theta^{-1})^2 = 0 = (\theta^1)^2$. There is
no triangle, so (\ref{2form_Delta}) reads $\Delta = 0$. But now there is a
biangle and (\ref{2form_zeta}) becomes
$\zeta = \theta^{-1} \theta^1 + \theta^1 \theta^{-1}$. Realizing $\A$ as the
algebra of functions on $\mathbb{Z}$, we recover the one-dimensional case
of the ``symmetric lattice'' treated in Ref.~\citen{DMH94}.

Corresponding differential calculi on $\mathbb{C}[x_1, \ldots, x_m]$ can be
built as skew tensor products of such calculi.
\end{example}

\begin{example}
Let $\A$ be the algebra of $\mathbb{C}$-valued functions on a discrete
group $G$. Then $G$ acts on itself by right translations which induce
automorphisms $R^\ast_g$ of $\A$ via $(R^\ast_g f)(g') = f(g'g)$.
Clearly, $\{ R_g^\ast \, | \, g \in G \}$ acts faithfully on $\A$.
Let $S$ be a finite subset of $G \setminus \{ e \}$.
Then $(G,S)$ is called a \emph{group lattice}.
According to the procedure of section~\ref{sec:dc_auto}, the set
of automorphisms $\{ R^\ast_s \, | \, s \in S \}$ determines a differential
calculus over $\A$ such that $\hat{\Omega}^1(\A) = \Omega^1(\A)$.
It is not in general inner at higher than first order.
Furthermore, in general, the maps $R_s^\ast$ do \emph{not}
extend to automorphisms of the corresponding differential calculus.
They do if $\mbox{ad}(S)S \subset S$ \cite{DMH02a}.
Every first-order group lattice differential calculus is simple.
Noncommutative geometry of group lattices has been elaborated in
Refs.~\citen{DMH02a,DMH02b}, to which we refer the reader for further
details.
\end{example}

\begin{example}
Let $\A$ be the algebra of $\mathbb{C}$-valued functions on the space
$G/H$ of right $H$-cosets of a group $G$ with respect to a subgroup $H$.
Then $G$ acts on $G/H$ by right translations
which induce automorphisms $R^\ast_g$ of $\A$ via $(R^\ast_g f)(Hg') = f(Hg'g)$.
Let $S$ be a finite subset of $G \setminus \{ e \}$. If certain conditions are satisfied,
the set of automorphisms $\{ R^\ast_s \, | \, s \in S \}$ determines a differential calculus
over $\A$, following the procedure of section~\ref{sec:dc_auto}.
Clearly, with the choice $H = \{ e \}$ we are back to our preceding example.
On the other hand, if $H = G$, then $G/H$ consists of a single element only.
In this case we have $R_g^\ast = \mathit{id}$ for all $g \in G$
and therefore not a faithful action.
If $H$ is a true subgroup of $G$ and if we can find a subgroup $G'$ of $G$
such that $R_{g'}^\ast \neq R_{g''}^\ast$ on $\A$ for all $g',g'' \in G'$ with
$g' \neq g''$, then a differential calculus can be built with each subset
$S \subset G' \setminus \{ e \}$. In Ref.~\citen{DMH02a} we also considered
coset differential calculi for which some of the $e_s$ vanish identically
on $\A$. In the present work, such cases are excluded.
\end{example}

\begin{example}
\label{ex:qplane}
The \emph{quantum plane} is the $\mathbb{C}$-algebra $\A$ generated
by elements $x,y$, subject to the relation $xy = q \, yx$ where $q \in \mathbb{C}$.
An element $f(x,y) \in \A$ is a polynomial in the variables $x,y$.
If $q \not\in \{\pm 1 \}$, the only automorphisms of $\A$ are given by
scalings of $x$ and $y$ \cite{Alev+Cham92}.
Let us choose $S = \{ \hat{1}, \hat{2} \} \subset \mathbb{Z}^2$, where
$\hat{1}=(1,0)$, $\hat{2}=(0,1)$, and corresponding automorphisms
\be
   (\phi_{\hat{1}}f)(x,y) = f(\alpha^{-1}x,\beta^{-1}y) \, , \qquad
   (\phi_{\hat{2}}f)(x,y) = f(\gamma^{-1}x,\delta^{-1}y) \, ,
   \label{qplane_auto}
\ee
with complex numbers $\alpha,\beta,\gamma,\delta \neq 0$. This extends
to a faithful action of the group $\mathbb{Z}^2$ (with addition as the
group composition) if $\ln\alpha \, \ln\delta \neq \ln\beta \, \ln\gamma$
(choosing an appropriate branch of the logarithm) and
$(\alpha,\beta) \neq (1,1) \neq (\gamma,\delta)$. $\{ e_s | s \in S \}$, and
then also $\{ \theta^s | s \in S \}$, is a left and right $\A$-module basis.
(\ref{d_e}) with (\ref{e_s}), using (\ref{cmr_thetaf}), leads to
\be
  \d x = A \, \theta^{\hat{1}} \, x
         + C \, \theta^{\hat{2}} \, x \, , \qquad
  \d y = B \, \theta^{\hat{1}} \, y
         + D \, \theta^{\hat{2}} \, y   \label{dxdy_theta_qplane}
\ee
where
\be
    A = (1 - \alpha)/t_{\hat{1}} \, , \quad
    B = (1 - \beta)/t_{\hat{1}} \, , \quad
    C = (1 - \gamma)/t_{\hat{2}} \, , \quad
    D = (1 - \delta)/t_{\hat{2}} \; .
\ee
Now we obtain
\be
  x \, \d x &=& ( \alpha \, A \, \theta^{\hat{1}}
                + \gamma \, C \, \theta^{\hat{2}} ) \, x^2  \, , \quad
  y \, \d x  = q^{-1} \, ( \beta \, A \, \theta^{\hat{1}}
                + \delta \, C \, \theta^{\hat{2}} ) \, x \, y \, ,
                \nonumber \\
  y \, \d y &=& ( \beta \, B \, \theta^{\hat{1}}
                + \delta \, D \, \theta^{\hat{2}} ) \, y^2 \, , \quad
  x \, \d y  = q \, ( \alpha B \, \theta^{\hat{1}}
                + \gamma \, D \, \theta^{\hat{2}} ) \, y \, x \; .
\ee
Under what conditions on the constants appearing in these equations can we
express the right hand sides in terms of $x,y,\d x,\d y$ ?
Let us extend the algebra $\A$ with additional generators $x^{-1}, y^{-1}$
subject to the relations $x x^{-1} = \mathbf{1} = x^{-1} x$ and
$y y^{-1} = \mathbf{1} = y^{-1} y$ with unit $\mathbf{1}$. The resulting algebra
$\hat{\A}$ is known as that of the \emph{quantum torus} and the maps (\ref{qplane_auto})
extend to it as automorphisms in an obvious way. Assuming $AD-BC \neq 0$, we find
\be
  \theta^{\hat{1}} = (AD-BC)^{-1} ( D \, \d x \, x^{-1} - C \, \d y \, y^{-1} )
   , \;
  \theta^{\hat{2}} = (AD-BC)^{-1} ( A \, \d y \, y^{-1} - B \, \d x \, x^{-1} ) , \quad
\ee
and the above commutation relations take the form
\be
  x \, \d x &=& (AD-BC)^{-1} \, [ ( \alpha \, AD - \gamma \, BC ) \, \d x
                + ( \gamma - \alpha) \, AC \, \d y \, y^{-1} \, x ] \, x  \, ,
                \nonumber \\
  y \, \d x &=& (AD-BC)^{-1} \, [ q^{-1} (\beta \, AD - \delta \, BC) \, \d x \, y
                + (\delta - \beta) \, AC \, \d y \, x ] \, , \nonumber \\
  y \, \d y &=& (AD-BC)^{-1} \, [ (\beta - \delta) \, BD \, \d x \, x^{-1} y
                + (\delta \, AD - \beta \, BC) \, \d y ] \, y \, , \nonumber \\
  x \, \d y &=& (AD-BC)^{-1} \, [ (\alpha - \gamma) \, BD \, \d x \, y
                + q \, (\gamma \, AD - \alpha \, BC) \, \d y \, x ] \; .
\ee
They reduce to relations on the quantum plane if the coefficients of all terms
containing $x^{-1}$ or $y^{-1}$ vanish, i.e.,
$( \gamma - \alpha) \, AC = 0 = (\beta - \delta) \, BD$. This leads to the
following three cases.  \\
a) $\delta = \beta$. Then $AC = 0$, so that $A=0$ or $C=0$, which means
$\alpha =1$ or $\gamma =1$. Let us choose $\gamma=1$. Then
\be
  x \, \d x &=& \alpha \, \d x \, x \, , \quad
  y \, \d x  = q^{-1} \beta \, \d x \, y  \nonumber \\
  y \, \d y &=& \beta \, \d y \, y \, , \quad
  x \, \d y  = q \, \d y \, x + (\beta - 1) \, \d x \, y \, .
    \label{q_plane_rels}
\ee
The alternative case $\alpha =1$ leads to the same relations, but with
$\alpha$ replaced by $\gamma$.
With the choice $\alpha = \beta = pq$ we obtain from (\ref{q_plane_rels})
the relations
\be
  x \, \d x &=& p q \, \d x \, x \, , \quad y \, \d x = p \, \d x \, y \nonumber \\
  y \, \d y &=& p q \, \d y \, y \, , \quad
  x \, \d y = q \, \d y \, x + (pq-1) \, \d x \, y  \; .
     \label{qplane_diff}
\ee
This calculus is covariant under the quantum group $GL_{p,q}(2)$ \cite{Schi91}
(see also Ref.~\citen{qplane}).  \\
b) $\gamma = \alpha$. This implies $BD = 0$.
The two solutions which emerge in this case are mapped to (\ref{q_plane_rels})
via $x \leftrightarrow y$, $q \mapsto q^{-1}, \alpha \mapsto \beta$,
and $\delta \mapsto \alpha$, respectively $\beta \mapsto \alpha$. \\
c) $\gamma \neq \alpha$ and $\delta \neq \beta$. Then $AC = 0 = BD$, so
that either $B = C = 0$ or $A = D = 0$. Let us consider the first case,
where $\beta = \gamma = 1$. Then
\be
   x \, \d x = \alpha \, \d x \, x \, , \quad
   y \, \d x = q^{-1} \, \d x \, y \, , \quad
   y \, \d y = \delta \, \d y \, y \, , \quad
   x \, \d y = q \, \d y \, x  \; .
\ee
If $A = D = 0$, so that $\alpha = \delta = 1$, we obtain the same
relations, but with $\alpha$ and $\delta$ replaced by $\gamma$
and $\beta$, respectively.
\vskip.1cm

In the case under consideration, there are only quadrangles so that
$\Delta(\theta^s) = 0 = \zeta$ and the associated
2-form relations (\ref{2formrel}) are
\be
  (\theta^{\hat{1}})^2 = (\theta^{\hat{2}})^2 = 0 \, , \qquad
   \theta^{\hat{1}} \, \theta^{\hat{2}} + \theta^{\hat{2}}
   \, \theta^{\hat{1}} = 0  \; .
\ee
As a consequence, $\d \vartheta = [\vartheta,\vartheta] = 2 \, \vartheta^2 = 0$.
The whole differential calculus is inner: $\d \omega = [\vartheta,\omega]$
for all $\omega \in \hat{\Omega}(\A)$.
The first order calculus is simple if $q$ is not a root of $\beta,\delta$,
and $q^{-1}$ not a root of $\alpha,\gamma$.
Furthermore, using the criterion (\ref{phi_diff_criterion}), it follows
that the automorphisms $\phi_s$ and their inverses are differentiable
with respect to each of the above differential calculi over the extended
algebra $\hat{\A}$. Application to (\ref{dxdy_theta_qplane})
leads to $\phi_s(\theta^{\hat{1}}) \, x = \theta^{\hat{1}} \, x$ and
$\phi_s(\theta^{\hat{2}}) \, y = \theta^{\hat{2}} \, y$. In the extended
algebra, this is
\be
        \phi_s (\theta^{s'}) = \theta^{s'} \; .
\ee
It follows that, with these relations, the maps $\phi_s$ are also
differentiable with respect to each of the above (generalized) differential
calculi over $\A$.
\end{example}

\begin{example}
The \emph{Heisenberg algebra} is generated by two elements $x,y$, subject
to the relation $[x,y] = h \, \mathbf{1}$ with $h \in \mathbb{C}$. Then
\be
  (\phi_{\hat{1}}f)(x,y) = f(x+a \mathbf{1},y) \, , \qquad
  (\phi_{\hat{2}}f)(x,y) = f(x,y+b \mathbf{1}) \, ,
\ee
are automorphisms, where $a,b \in \mathbb{C} \setminus \{0\}$. This yields a
faithful action of $G = \mathbb{Z}^2$. Choosing
$t_{\hat{1}} = a$ and $t_{\hat{2}} = b$, we obtain
\be
  \d x = \theta^{\hat{1}} \, , \qquad
  \d y = \theta^{\hat{2}} \, ,
\ee
and the commutation relations
\be
   [\d x, x] = a \, \d x \, , \qquad
   [\d x,y] = 0 \, , \qquad
   [\d y,x] = 0 \, , \qquad
   [\d y,y] = b \, \d y \; .
\ee
The automorphisms $\phi_s$ preserve these relations and are therefore
differentiable. They satisfy $\phi_s(\theta^{s'}) = \theta^{s'}$.
In the limit $h \to 0$ we obtain a familiar differential calculus
which supplies $\mathbb{R}^2$ (or $\mathbb{Z}^2$) with a quadratic
lattice digraph structure \cite{DMHS93}.
\end{example}

\begin{example}
The algebra $\A$ of the {\em $h$-deformed plane} is generated
by elements $x,y$, with the relation
\be
     [x, y] = h \, y^2   \label{hplane}
\ee
where $h \in \mathbb{C}$. Extending $\A$ with an element $y^{-1}$,
imposing the relations $y y^{-1} = \mathbf{1} = y^{-1} y$,
(\ref{hplane}) becomes equivalent to the correspondingly extended
Heisenberg algebra with $[y^{-1},x]= h \, \mathbf{1}$.
Two automorphisms of the (extended) $h$-plane are given by
\be
  (\phi_1 f)(x,y) = f(x + p y,y) \, , \qquad
  (\phi_2 f)(x,y) = f(r^{-1}x,r^{-1}y) \, ,
\ee
with $p,r \in \mathbb{C} \setminus \{0\}$. They generate a faithful action of
$\mathbb{Z}^2$ on $\A$. The 1-forms $\theta^{\hat{1}}, \, \theta^{\hat{2}}$
of the associated differential calculus then satisfy
\be
  f(x,y) \, \theta^{\hat{1}} = \theta^{\hat{1}} \, f(x-p y,y) \, , \qquad
  f(x,y) \, \theta^{\hat{2}} = \theta^{\hat{2}} \, f(r x,r y) \; .
\ee
The vector fields given by (\ref{e_s}) act on $f \in \A$ as follows,
\be
 (e_1 f)(x,y) = [ f(x+py,y) - f(x,y) ]/t_1 \, , \quad
 (e_2 f)(x,y) = [ f(r^{-1}x,r^{-1}y) - f(x,y) ]/t_2 \, , \quad
\ee
and (\ref{d_e}) yields
\be
  \d x = \frac{p}{t_1} \, \theta^{\hat{1}} \, y + \frac{1-r}{t_2}
         \, \theta^{\hat{2}} \, x   \, , \qquad
  \d y = \frac{1-r}{t_2} \, \theta^{\hat{2}} \, y  \; .
  \label{dxdy_theta_hplane}
\ee
Now we obtain
\be
  && x \, \d x = \frac{p}{t_1} \, \theta^{\hat{1}} \, [ y \, x + (h-p) \, y^2 ]
       + \frac{r-r^2}{t_2} \, \theta^{\hat{2}} \, x^2 \, , \quad
  x \, \d y = \frac{r-r^2}{t_2} \, \theta^{\hat{2}} \, (y \, x + h \, y^2 ) \, ,
                 \nonumber \\
  && y \, \d x = \frac{p}{t_1} \, \theta^{\hat{1}} \, y^2
       + \frac{r-r^2}{t_2} \, \theta^{\hat{2}} \, ( x \, y -h \, y^2 ) \, , \quad
  y \, \d y = \frac{r-r^2}{t_2} \, \theta^{\hat{2}} \, y^2 \; .
  \label{hplane_xdx}
\ee
In order to replace the $\theta^s$ by expressions involving $\d x$ and $\d y$,
we return to the extended algebra. Then
\be
   \theta^{\hat{1}} = \frac{t_1}{p} \, ( \d x - \d y \, y^{-1} x ) \, y^{-1}
                      \, , \qquad
   \theta^{\hat{2}} = \frac{t_2}{1-r} \, \d y \, y^{-1} \; .
   \label{theta_dxdy_hplane}
\ee
Substitution into the above commutation relations leads to
\be
  \lbrack x , \d x ] &=& h' \, ( \d y \, ( x + h \, y) - \d x \, y )
                   + (r-1) \, \d y \, y^{-1} x^2 \, , \nonumber \\
  \lbrack y , \d x ] &=& - h \, \d y \, y  + (r-1) \, \d y \, x \, , \nonumber \\
  \lbrack y , \d y ] &=& (r-1) \, \d y \, y \, , \nonumber \\
  \lbrack x , \d y ] &=& r \, h \, \d y \, y + (r-1) \, \d y \, x
        \label{hplane_diff}
\ee
where we introduced $h' = p-h$.
The only way to get rid of the terms containing $y^{-1}$ is to take
the limit $r \to 1$. Then we recover a differential calculus which is
covariant under the quantum group $GL_{h,h'}(2)$ \cite{Agha93}.
In this limit $\theta^{\hat{2}}$ commutes with all $f \in \A$.
In particular, the resulting (first order) differential calculus
is not simple. In order for $e_2$ to have a limit,
we should choose $t_2 = 1-r$. $e_2$ then tends to an outer
derivation on $\A$ as $r \to 1$.

 From the properties of the group $\mathbb{Z}^2$, we deduce
$\Delta(\theta^s) = \zeta = 0$ and
\be
  (\theta^{\hat{1}})^2 = (\theta^{\hat{2}})^2 = 0 \, , \qquad
  \theta^{\hat{1}} \theta^{\hat{2}} + \theta^{\hat{2}} \theta^{\hat{1}} = 0 \; .
  \label{hplane_2form_rels}
\ee
 From these relations we obtain, in the limit $r \to 1$,
\be
   (\d x)^2 = h' \, \d x \, \d y \, , \qquad
   (\d y)^2 = 0 \, , \qquad
   \d x \, \d y + \d y \, \d x = 0
\ee
(see also Ref.~\citen{Agha93}).

Differentiability of the maps $\phi_s$ (with respect to the differential
calculus over the extended algebra) requires $\phi_s(\theta^{s'}) = \theta^{s'}$
via (\ref{theta_dxdy_hplane}). It follows that $\phi_s$ preserves the relations
(\ref{hplane_xdx}) and is therefore indeed differentiable. $\phi_s$
is then also differentiable with respect to the (generalized)
differential calculus over $\A$ obtained in the limit $r \to 1$.
\end{example}

\begin{example}
Let $\A$ be the complex algebra freely
generated by $x,y,x^{-1},y^{-1}$, subject to
$x x^{-1} = \mathbf{1} = x^{-1} x$ and
$y y^{-1} = \mathbf{1} = y^{-1} y$.
Furthermore, let $q$ be a primitive root of unity of degree three,
so that $q^3=1$ and $1+q+q^2=0$. Then
\be
  (\phi_1 f)(x,y) = f(q x,q^2 y) \, , \qquad
  (\phi_2 f)(x,y) = (\phi_1^2 f)(x,y) = f(q^2 x,q y)
\ee
determines a faithful action of $\mathbb{Z}_3$ on $\A$ by automorphisms.
In the associated differential calculus we have
\be
  \theta^1 \, f(x,y) = f(q x,q^2 y) \, \theta^1 \, , \qquad
  \theta^2 \, f(x,y) = f(q^2 x,q y) \, \theta^2
\ee
and, using (\ref{d_e}) with (\ref{e_s}) and $t_1=t_2=q-1$,
\be
  \d x = x \, \theta^1 - q^2 \, x \, \theta^2 \, , \qquad
  \d y = -q^2 \, y \, \theta^1 + y \, \theta^2 \; .
\ee
This implies
\be
  \theta^1 = {1 \over 1-q} (x^{-1} \, \d x + q^2 \, y^{-1} \, \d y) \, , \quad
  \theta^2 = {1 \over 1-q} (q^2 \, x^{-1} \, \d x + y^{-1} \, \d y)
  \label{q^3=1_theta}
\ee
and
\be
   \d x \, x &=& - x \, \d x + x^2 \, y^{-1} \, \d y \, , \quad
   \d x \, y  = -x \, \d y \, , \nonumber \\
   \d y \, y &=& -y \, \d y + y^2 \, x^{-1} \, \d x \, , \quad
   \d y \, x  = -y \, \d x   \,.   \label{q^3=1_xdx}
\ee
 Using these relations and the Leibniz rule, we find
\be
  \d(x^2) = x^2 y^{-1} \, \d y \, , \qquad
  \d(y^2) = y^2 x^{-1} \, \d x \, ,
\ee
and
\be
  \d (x^3) = \d(y^3) = \d(x y) = \d (y x) = 0 \; .
\ee
As a consequence,
\be
   x^3 = c_1 \, , \quad
   y^3 = c_2 \, , \quad
   x y = c_3 \, , \quad
   y x = c_4 \, ,
\ee
are constants of the calculus, i.e., $\phi_s(c_i) = c_i$, $s = 1,2$ and $i = 1,2,3,4$.
Furthermore, we find
\be
   \d (x^2) = x^3 (y x)^{-1} \, \d y = c_1 \, c_4^{-1} \, \d y \, , \qquad
   \d (y^2) = y^3 (x y)^{-1} \, \d x = c_2 \, c_3^{-1} \, \d x \, ,
\ee
and thus
\be
    \d ( x^2 - c_1 c_4^{-1} y ) = 0 \, , \qquad
    \d ( y^2 - c_2 c_3^{-1} x ) = 0 \, ,
\ee
so we can set $x^2 - c_1 c_4^{-1} y = c_5$ and $y^2 - c_2 c_3^{-1} x = c_6$.
Multiplying the first equation with $x$ and the second with $y$ from the right
leads to $c_5 = 0 = c_6$. Hence
\be
    x^2 = c_1 \, c_4^{-1} \, y \, , \qquad
    y^2 = c_2 \, c_3^{-1} \, x \; .
\ee
It follows that every $f \in \A$ can be written as a linear combination of
$\mathbf{1},x,y$, where the coefficients are constants with respect to the
differential calculus.

If we divide $\A$ by the ideal generated by
$c_1-\mathbf{1},c_2-\mathbf{1},c_3-\mathbf{1},c_4-\mathbf{1}$, we obtain an
algebra $\A'$ generated by $x$ with the relation $x^3=\mathbf{1}$. The elements
\be
  e^0 = {1 \over3} \, (\mathbf{1}+x+x^2) \, , \quad
  e^1 = {1 \over3} \, (\mathbf{1}+q^2 x+q x^2) \, , \quad
  e^2 = {1 \over3} \, (\mathbf{1}+q x+q^2 x^2)
\ee
of $\A'$ are then primitive idempotents and satisfy $e^0+e^1+e^2 = \mathbf{1}$.
As a consequence, $\A'$ becomes the algebra of functions on $\mathbb{Z}_3$
and the above differential calculus is the universal one.

The maps $\phi_s$ obviously preserve the relations (\ref{q^3=1_xdx}) in
the sense of (\ref{phi_diff_criterion}) and are therefore differentiable.
(\ref{q^3=1_theta}) then implies $\phi_s(\theta^{s'}) = \theta^{s'}$.
\end{example}

\begin{example}
\label{ex:matrix}
Let $\A$ be a noncommutative matrix algebra and $G$ a group
which acts faithfully on $\A$ by inner automorphisms
$\{ \phi_g \, | \, g \in G \}$, so that
$\phi_g(f) = U_g f U_g^{-1}$ with an invertible matrix $U_g$.
Each set $S \subset G \setminus \{ e \}$ then determines a
differential calculus.
If we have 1-parameter groups $U_s = e^{t_s \lambda_s}$ with matrices
$\lambda_s$ and $t_s \in \mathbb{R}$, in the limit $t_s \to 0$
we obtain $\theta^s f = f \theta^s$ and
\be
   \lim_{t_s \to 0} e_s f = \lim_{t_s \to 0} {\phi_s(f) - f \over t_s}
                          = [\lambda_s,f]
\ee
so that each $e_s$ tends to a derivation, a special case of
(\ref{twist_deriv}).
Assuming $U_s^{-1} \in \A$, (\ref{cmr_thetaf})
implies that $U_s^{-1} \theta^s$, $s \in S$, are 1-forms which commute
with all $f \in \A$ (see also Lemma~\ref{lemma:simple}).
Then we can modify $\vartheta$ as follows,
\be
     \vartheta'
   = \vartheta - \sum_{s \in S} {1 \over t_s} U_s^{-1} \theta^s
   = \sum_{s \in S}  {\mathbf{1} - U_s^{-1} \over t_s} \theta^s
\ee
while keeping $\d f = [\vartheta',f]$. In contrast to $\vartheta$, the
1-form $\vartheta'$ has a limit:
\be
  \lim_{\{t_s\} \to 0} \vartheta' = \sum_{s \in S} \lambda_s \, \theta^s \; .
\ee
The differential calculi obtained in this limit have been explored
in Ref.~\citen{Dima+Mado96}. They are not reached directly by the
construction presented in the main part of this section and demand
a suitable generalization which is the subject of the following section.
\end{example}

\section{Differential calculi associated with twisted inner derivations}
\label{sec:general}
\setcounter{equation}{0}
Example~\ref{ex:matrix} (see also Ref.~\citen{Dima+Mado96}) demonstrates
that the construction of differential calculi associated with automorphisms
presented in section~\ref{sec:dc_auto} does not exhaust the possibilities
offered by (\ref{theta_f_phi}) and suggests a suitable generalization.
Let $\phi_s$ be automorphisms of $\A$ and $\lambda_s \in \A$.
We introduce ``twisted inner derivations" by
\be
  e_s(f) = \lambda_s \, \phi_s(f) - f \, \lambda_s
   \qquad \forall f \in \A        \label{twist_deriv}
\ee
with $\lambda_s \in \A$. They satisfy
\be
   e_s(f f') =  e_s(f) \, \phi_s(f') + f \, e_s(f')
                \qquad  \forall f,f' \in \A \; .
\ee
Using the relations (\ref{ef1}) and (\ref{ef2}), we obtain an $\A$-bimodule
$\X$ and via (\ref{dual1})-(\ref{dual3}) a dual module $\hat{\Omega}^1(\A)$.
If $\{ e_s | s \in S \}$ is a basis of $\X$, let $\{ \theta^s | s \in S \}$
be the dual basis. Then (\ref{theta_f_phi}) holds and
\be
  \d f = [\vartheta,f] = \sum_{s \in S} e_s(f) \, \theta^s  \label{df_twist}
\ee
with
\be
     \vartheta = \sum_s \lambda_s \, \theta^s  \label{vartheta_lambda}
\ee
defines a first order differential calculus $(\hat{\Omega}^1(\A),\d)$.
All formulas and results of section~\ref{sec:dc_auto}, with the exception of
the subsections, remain valid after replacing (\ref{e_s}) by (\ref{twist_deriv}).
We only have to exchange (\ref{vartheta_t}) with (\ref{vartheta_lambda}) and
note that now $c$ is a \emph{constant} of the differential
calculus iff $c \, \lambda_s = \lambda_s \, \phi_s(c)$ for all $s \in S$.
\vskip.1cm

\begin{example}
The defining relations of the quantum plane differential calculus
of example~\ref{ex:qplane} can be presented in various ways if we
relax (\ref{e_s}) and allow twisted derivations, while keeping the
structure (\ref{theta_f_phi}). Let us look more closely at
the covariant calculus (\ref{qplane_diff}) where
$\theta^{\hat{1}} \, (x,y) = r^{-1} \, (x,y) \, \theta^{\hat{1}}$
and $\theta^{\hat{2}} \, (x,y) = (x, r^{-1} y) \, \theta^{\hat{2}}$
with $r=pq$. Let $\theta^1 := x^k y^l \, \theta^{\hat{1}}$ and
$\theta^2 := x^m y^n \, \theta^{\hat{2}}$, so that
$\lambda_1 = y^{-l}x^{-k}/t_{\hat{1}}$,
$\lambda_2 = y^{-n} x^{-m}/t_{\hat{2}}$, assuming that $x,y$ are invertible.
Then $\theta^1 \, (x,y) = r^{-1} \, (q^{-l} x, q^k y) \, \theta^1(k,l)$,
$\theta^2 \, (x,y) = (q^{-n} x, r q^m y) \, \theta^2$.
In particular, if $p=q$ we can choose $k=2, l=-2, m=2,n=0$, and the
1-forms $\theta^1$ and $\theta^2$ commute with all elements of $\A$
(see also Ref.~\citen{Dima+Mado96}). In the latter case, (\ref{twist_deriv})
become ordinary inner derivations and the automorphism structure is hidden.
\end{example}

In the first subsection we explore the structure of 2-forms.
The second subsection reveals the above structure in a bicovariant
differential calculus on the quantum group $GL_{p,q}(2)$.

\subsection{The structure of 2-forms}
Using (\ref{zeta}) and (\ref{vartheta_lambda}), we find
\be
 \zeta = \sum_{s \in S} \lambda_s \, \d \theta^s
       - \sum_{s,s' \in S} \lambda_{s'} \lambda_s \, \theta^s \theta^{s'}
       = - \sum_{s \in S} \lambda_s \, \Delta(\theta^s)
         + \sum_{s,s' \in S} \lambda_s \, \phi_s(\lambda_{s'}) \, \theta^s \theta^{s'} \; .
\ee
Commuting $f \in \A$ from the right towards the left of each term in this
formula, using (\ref{theta_f_phi}) and (\ref{zeta_f}), leads to
\be
   \sum_{s,s' \in S} \Big( f \, \lambda_s \, \phi_s(\lambda_{s'})
      - \lambda_s \, \phi_s(\lambda_{s'}) \, \phi_{ss'}(f) \Big) \, \theta^s \theta^{s'}
    = \sum_{s \in S} \Big( f \, \lambda_s - \lambda_s \, \phi_s(f) \Big) \, \Delta(\theta^s) \; .
    \label{2-form-relations}
\ee
Its elaboration in concrete examples yields the 2-form relations
and determines the map $\Delta$.
\vskip.1cm

Let $\Xi$ be a subset of $S \times S$ such that
$\{ \theta^s \theta^{s'} | (s,s') \in \Xi \}$
is a left $\A$-module basis of $\hat{\Omega}^2(\A)$. Writing
\be
    \zeta = \sum_{(s,s') \in \Xi} \zeta_{s,s'} \, \theta^s \theta^{s'} \, ,
\ee
(\ref{zeta_f}) demands
\be
    f \, \zeta_{s,s'} = \phi_{ss'}(f) \, \zeta_{s,s'}
    \qquad \forall f \in \A, \; (s,s') \in \Xi \; .
\ee
\vskip.1cm

\begin{example}
Let us consider the Heisenberg algebra generated by $x,y$ with $[x,y] = \mathbf{1}$.
We choose $\phi_s = \mbox{id}$, $s=1,2$, and $\lambda_1 = -y$, $\lambda_2 = x$.
Then $e_s f = [\lambda_s , f]$, $\theta^1 = \d x$, $\theta^2 = \d y$,
$\theta^s \, f = f \, \theta^s$, and $\vartheta = x \, \d y - y \, \d x$.
Evaluation of (\ref{2-form-relations}) leads to the 2-form relations
$(\theta^s)^2 = 0$, $\theta^2 \theta^1 = - \theta^1 \theta^2$, and
moreover requires $\Delta(\theta^s) = 0$. Hence we may choose $\Xi = \{ (1,2) \}$.
Furthermore, (\ref{zeta}) shows that $\zeta = \theta^1 \theta^2$.

If we add one more derivation with $\lambda_3 = y x$, there is a
third 1-form $\theta^3$. Inspection of (\ref{2-form-relations}) then yields
$\Delta(\theta^1) = - \theta^1 \theta^3$, $\Delta(\theta^2) = \theta^2 \theta^3$,
$\Delta(\theta^3) = - \theta^1 \theta^2 - \theta^2 \theta^1$,
and the 2-form relations
$(\theta^1)^2 = (\theta^2)^2 = (\theta^3)^2 = 0$,
$\theta^3 \theta^1 = - \theta^1 \theta^3$,
$\theta^3 \theta^2 = - \theta^2 \theta^3$.
Hence we can choose $\Xi = \{(1,2),(2,1),(1,3),(2,3) \}$.
Furthermore, $\zeta = - \theta^2 \theta^1$.
\end{example}

\subsection{A bicovariant calculus on $GL_{p,q}(2)$}
\label{sec:GLpq(2)}
The quantum group $GL_{p,q}(2)$ is given by the algebra $\A$ generated
by elements $a,b,c,d$ with relations
\be
 &&a \, b = p \, b \, a \, , \quad
   a \, c = q \, c \, a \, , \quad
   b \, c = (q/p) \, c \, b \, , \quad
   b \, d = q \, d \, b \, , \quad
   c \, d = p \, d \, c \, , \nonumber \\
 &&a \, d = d \, a + (p - q^{-1}) \, b \, c
\ee
and carries a Hopf algebra structure.\footnote{The existence of an antipode
requires that the quantum determinant $\cal D$ (see below) is invertible. In
most of what follows we will not need this condition, however.}
Its automorphisms are given by scalings
$ (a,b,c,d) \mapsto (\alpha a, \beta b, \gamma c, \delta d) $
with $\alpha, \beta, \gamma, \delta \in \mathbb{C} \setminus \{0\}$ such that
$\alpha \delta = \beta \gamma$.\footnote{If $p=q$, in which
case the quantum group is denoted as $GL_q(2)$,
there is an additional automorphism which interchanges $b$ and $c$,
and leaves $a,d$ fixed \cite{Alev+Cham92}.}
Of course, we can use these automorphisms to construct differential
calculi on $GL_{p,q}(2)$ according to the prescription of section~\ref{sec:dc_auto}.
An interesting question is whether the distinguished {\em bicovariant}
differential calculi on $GL_{p,q}(2)$ \cite{MH92} are of this form or at least
exhibit the more general structure described in the beginning of this section.
Let us look more closely at one of them which is determined in terms of
generalized Maurer-Cartan 1-forms $\tth^i$, $i=1,\ldots,4$, by the relations
\be
&&\tth^1 a = r \, a \, \tth^1 + (r-1) \, b \, \tth^3 \, , \quad
  \tth^2 a = q \, a \, \tth^2 + p^{-1} (r-1) \, b \, \tth^4 \, , \quad
  \tth^3 a = p \, a \, \tth^3 \, , \quad
  \tth^4 a = a \, \tth^4 \, ,  \nonumber \\
&&\tth^1 b = b \, \tth^1 + (r-1) \, a \, \tth^2 + r^{-1} (r-1)^2 \, b \, \tth^4 \, , \quad
  \tth^2 b = q \, b \, \tth^2 \, , \nonumber \\
&&\tth^3 b = p \, b \, \tth^3 + q^{-1} (r-1) \, a \, \tth^4 \, , \quad
  \tth^4 b = r \, b \, \tth^4 \, ,  \nonumber \\
&&\tth^1 c = r \, c \, \tth^1 + (r-1) \, d \, \tth^3 \, , \quad
  \tth^2 c = q \, c \, \tth^2 + p^{-1} (r-1) \, d \, \tth^4 \, , \quad
  \tth^3 c = p \, c \, \tth^3 \, , \quad
  \tth^4 c = c \, \tth^4 \, ,  \nonumber \\
&&\tth^1 d = d \, \tth^1 + (r-1) \, c \, \tth^2 + r^{-1} (r-1)^2 \, d \, \tth^4 \, , \quad
  \tth^2 d = q \, d \, \tth^2 \, , \nonumber \\
&&\tth^3 d = p \, d \, \tth^3 + q^{-1} (r-1) \, c \, \tth^4 \, , \quad
  \tth^4 d = r \, d \, \tth^4       \label{GLpq_diff}
\ee
where $r = pq$ (see \cite{MH92}, example~1 of section 6). The action of the
exterior derivative $\d$ on $GL_{p,q}(2)$ is determined by
\be
   \d a = a \, \tth^1 + b \, \tth^3 \, , \quad
   \d b = a \, \tth^3 + b \, \tth^4 \, , \quad
   \d c = c \, \tth^1 + d \, \tth^3 \, , \quad
   \d d = c \, \tth^2 + d \, \tth^4 \; .
\ee
Let us introduce
\be
   \theta^4 = k_4 \, {\cal D}^P \, b^N \, c^{M} \, \tth^4
\ee
with $k_4 \in \mathbb{C} \setminus \{ 0 \}$, ${\cal D} = a \, d - p \, b \, c$
and arbitrary non-negative integers $M,N,P$.\footnote{Note that
${\cal D} \, a = a \, {\cal D}$, ${\cal D} \, b = (p/q) \, b \, {\cal D}$,
${\cal D} \, c = (q/p) \, c \, {\cal D}$, ${\cal D} \, d = d \, {\cal D}$.
The assumption of non-negative exponents can be relaxed if the corresponding
elements of $\A$ are assumed to be invertible.}
Then (\ref{theta_f_phi}) holds for $\theta^4$ with the automorphism $\phi_4$
which acts as follows on the generators:
\be
  && \phi_4(a) = p^{-N} \, q^{-M} \, a \, , \quad
     \phi_4(b) = r \, (p/q)^{M+P} \, b \, , \nonumber \\
  && \phi_4(c) = (q/p)^{N+P} \, c \, , \quad
     \phi_4(d) = p^{M+1} \, q^{N+1} \, d \; .
\ee
Moreover,
\be
   \theta^2 = k_2 \, {\cal D}^Q \, b^K \, c^L \, ( c \, \tth^2 + d \, \tth^4 )
\ee
with $k_2 \in \mathbb{C} \setminus \{ 0 \}$ and non-negative integers $Q,K,L$
also satisfies (\ref{theta_f_phi}) with
\be
  && \phi_2(a) = p^{-K} \, q^{-L} \, a \, , \quad
     \phi_2(b) = p^{Q+L+1} \, q^{-Q-L} \, b \, , \nonumber \\
  && \phi_2(c) = p^{-Q-K} \, q^{Q+K+1} \, c \, , \quad
     \phi_2(d) = p^{L+1} \, q^{K+1} \, d \; .
\ee
Furthermore,
\be
   \theta^3 = k_3 \, {\cal D}^R \, b^S \, c^T \, ( b \, \tth^3 -r^{-1} \, a \, \tth^4 )
\ee
with $k_3 \in \mathbb{C} \setminus \{ 0 \}$ and non-negative integers $R,S,T$
satisfies (\ref{theta_f_phi}) with
\be
  && \phi_3(a) = p^{-S} \, q^{-T} \, a \, , \quad
     \phi_3(b) = p^{R+T+1} \, q^{-R-T} \, b \, , \nonumber \\
  && \phi_3(c) = p^{-R-S} \, q^{R+S+1} \, c \, , \quad
     \phi_3(d) = p^{T+1} \, q^{S+1} \, d \; .
\ee
Finally,
\be
   \theta^1 = k_1 \, {\cal D}^U \, b^V \, c^W \, ( b \, c \, \tth^1
   - p^{-1} \, a \, c \, \tth^2 + b \, d \, \tth^3 - p^{-1} \, a \, d \, \tth^4 )
\ee
with $k_1 \in \mathbb{C} \setminus \{ 0 \}$ and non-negative integers $U,V,W$
satisfies (\ref{theta_f_phi}) with
\be
  && \phi_1(a) = p^{-V} \, q^{-W} \, a \, , \quad
     \phi_1(b) = (p/q)^{U+W+1} \, b \, , \nonumber \\
  && \phi_1(c) = q^2 (q/p)^{U+V} \, c \, , \quad
     \phi_1(d) = p^{W+1} \, q^{V+1} \, d \; .
\ee
Hence, in terms of the new 1-forms $\theta^s$ the commutation relations
of the above bicovariant differential calculus on $GL_{p,q}(2)$ indeed
take the form (\ref{theta_f_phi}) and thus have the ``automorphism structure''
we were looking for. This is quite surprising and raises the question to
what extent this result generalizes to other bicovariant differential
calculi on quantum groups (see \cite{dc_qg}, in particular).
\vskip.1cm

In order to express the Maurer-Cartan 1-forms $\tth^s$ as ordinary
1-forms, we need $\cal D$ to be invertible:
\be
 \left(\begin{array}{cc} \tth^1 & \tth^2 \\
                         \tth^3 & \tth^4
                 \end{array} \right)
 = {\cal D}^{-1}\, \left(\begin{array}{cc}
     d \,\mbox{d} a - q^{-1} \, b \, \mbox{d} c &
     d \,\mbox{d} b - q^{-1} \, b \, \mbox{d} d \\
     -q \, c \, \mbox{d} a + a \, \mbox{d} c &
     -q \, c \, \mbox{d} b + a \, \mbox{d} d
                 \end{array} \right)    \; .  \label{tth_d}
\ee
Let us write $(a_s) = (a,b,c,d)$ and $\phi_s (a_{s'}) = \alpha_{s,s'} \, a_{s'}$
where $\alpha_{s,s'} \in \mathbb{C}$. Using (\ref{tth_d}) we find that
\be
    \phi_s(\tth^1) = \tth^1 \, , \quad
    \phi_s(\tth^2) = (\alpha_{s,2}/\alpha_{s,1}) \, \tth^2 \, , \quad
    \phi_s(\tth^3) = (\alpha_{s,3}/\alpha_{s,4}) \, \tth^3 \, , \quad
    \phi_s(\tth^4) = \tth^4 \, ,
\ee
must hold for $\phi_s$ to be differentiable.\footnote{Note that
$\alpha_{s,1} \, \alpha_{s,4} = \alpha_{s,2} \, \alpha_{s,3}$.}
The maps $\phi_s$ then preserve the relations (\ref{GLpq_diff})
and are indeed differentiable. The action of $\phi_s$ on the $\theta^s$
is now easily obtained and also turns out to be homogeneous, i.e., it is
given by scalings with complex numbers.
The calculus is inner at first order and simple, but the 1-form
\be
    \vartheta = (r-1)^{-1} \, ( \tth^1 + r^{-1} \, \tth^4 )
\ee
is \emph{not} given by the formula (\ref{vartheta_t}).
\vskip.1cm

To be more explicit, let us choose
\be
 &&\theta^1 = (r-1)^{-1} \, c^{-1} b^{-1}(bc \, \tilde{\theta}^1
   - p^{-1} ac \, \tilde{\theta}^2
   + bd \, \tilde{\theta}^3 - p^{-1} ad \, \tilde{\theta}^4) \, ,
                                            \nonumber \\
 &&\theta^2 = q \, (r-1)^{-1} \, c^{-1} b^{-1} (c \, \tilde{\theta}^2
   + d \, \tilde{\theta}^4) \, ,            \nonumber \\
 &&\theta^3 = -[p (r-1)]^{-1} \, c^{-1} b^{-1} (b \, \tilde{\theta}^3
   - r^{-1} a \, \tilde{\theta}^4) \, ,     \nonumber \\
 &&\theta^4 = -[p (r-1)]^{-1} \, c^{-1} b^{-1} {\cal D} \, \tilde{\theta}^4 \, ,
     \label{GLpq(2)_thetas}
\ee
assuming that $b$ and $c$ are invertible. Then we obtain
\be
    \vartheta = \theta^1 + a \, \theta^2 + d \, \theta^3 + \theta^4
\ee
and thus
\be
   e_1(f) = \phi_1(f) -f \, , \;
   e_2(f) = a \, \phi_2(f) - f \, a \, , \;
   e_3(f) = d \, \phi_3(f) - f \, d \, , \;
   e_4(f) = \phi_4(f) - f \, , \quad
\ee
where the action of the automorphisms $\phi_s$ on the quantum group algebra
is determined by
\be
     (\alpha_{s,s'}) = \left(\begin{array}{cccc}
       r & 1 & r & 1 \\
       r & q & p & 1 \\
       r & q & p & 1 \\
       r & r & 1 & 1 \end{array}\right) \, .  \label{phi_alpha}
\ee
Furthermore,
\be
   \phi_s(\theta^1) = \theta^1 \, , \quad
   \phi_s(\theta^2) = r^{-1} \, \theta^2 \, , \quad
   \phi_s(\theta^3) = \theta^3 \, , \quad
   \phi_s(\theta^4) = \theta^4 \, ,  \label{GLpq(2)_phithetas}
\ee
which implies $\phi_s(\vartheta) = \vartheta$, in accordance with
Lemma~\ref{lemma:phi}.
\vskip.1cm

The bicovariant differential calculus extends to higher orders
in such a way that
\be
    \d \omega = [ \vartheta , \omega ]
\ee
for an arbitrary form $\omega$.\footnote{The generalized wedge product
has been obtained in Ref.~\citen{MH92}, see (6.27) therein.}
Hence $\Delta(\omega) = 0$ and $\zeta = \vartheta^2$.

\section{Connections}
\label{sec:conn}
\setcounter{equation}{0}
A {\em connection} on a left $\A$-module $\E$ is a linear map
$\nabla : \E \rightarrow \hat{\Omega}^1(\A) \oA \E$ such that
\be
   \nabla (f \, E) = \d f \oA E + f \, \nabla(E)  \qquad \forall E \in \E
   \label{nafE}
\ee
with respect to a (possibly generalized) differential calculus $(\hat{\Omega}(\A),\d)$.
It extends to a map
$\na : \hat{\Omega}(\A) \oA \E \rightarrow \hat{\Omega}(\A) \oA \E$ via
\be
   \na(\omega \oA E)
 = \d \omega \oA E + (-1)^r \omega \, \na E  \qquad
   \forall \omega \in \hat{\Omega}^r(\A), \, E \in \E  \, .
      \label{nabla_ext}
\ee
Its {\em curvature} is the left $\A$-module homomorphism
$\R : \E \rightarrow \hat{\Omega}^2(\A) \oA \E$ given by
\be
   \R(E) = - \na^2 E
\ee
which extends to a map $\hat{\Omega}(\A) \oA \E \rightarrow \hat{\Omega}(\A) \oA \E$
with the property
\be
    \R(\omega \oA E) = \omega \, \R(E)
    \qquad
    \forall \omega \in \hat{\Omega}(\A), \, E \in \E  \; .
\ee

\begin{lemma}
\label{lemma:connE}
Let $\d : \A \rightarrow \hat{\Omega}^1(\A)$ be inner.
Every connection on $\E$ is then of the form
\be
  \nabla(E) = \vartheta \oA E - \V(E)  \qquad \forall E \in \E
              \label{na_E}
\ee
where $\V : \E \rightarrow \hat{\Omega}^1 \oA \E$ satisfies
\be
    \V( f \, E ) = f \, \V(E)  \qquad \forall f \in \A \; .
    \label{V_left_tensor_prop}
\ee
Conversely, every linear map $\V$ with this property defines a
connection via the above formula.
\end{lemma}
{\bf Proof:} This is easily verified using (\ref{d_inner})
(see also Ref.~\citen{DMH02a}).
\hfill $\blacksquare$
\vskip.2cm

Now let $(\hat{\Omega}(\A),\d)$ be a (generalized) differential calculus
which is inner at first order and associated with automorphisms
$\{ \phi_s \, | \, s \in S \}$ of $\A$ such that (\ref{theta_f_phi})
holds. Writing
\be
    \V = \sum_{s \in S} \theta^s \oA \V_s  \label{VVs}
\ee
with parallel transport operators $\V_s$ (``in the $s$-direction''),
we find
\be
    \V_s(f \, E) = \phi_s^{-1}(f) \, \V_s(E)  \; .
    \label{VsfE}
\ee

\subsection{Linear connections}
If $\E = \hat{\Omega}^1(\A)$, $\nabla$ is called a \emph{linear connection}.
The {\em torsion} of $\nabla$ is the left $\A$-module
homomorphism $\Theta : \hat{\Omega}^1(\A) \rightarrow \hat{\Omega}^2(\A)$ defined by
\be
   \Theta(\alpha)
 = \d \alpha - \pi \circ \na \alpha
   \qquad   \forall \alpha \in \hat{\Omega}^1(\A)
\ee
where $\pi$ is the canonical projection
$\hat{\Omega}^1(\A) \oA \hat{\Omega}^1(\A) \rightarrow \hat{\Omega}^2(\A)$. It extends
to a map $\Theta : \hat{\Omega}(\A) \oA \hat{\Omega}^1(\A) \rightarrow \hat{\Omega}(\A)$ via
\be
   \Theta = \d \circ \pi - \pi \circ \na
\ee
with the property
\be
  \Theta (\omega \oA \alpha) = (-1)^r \, \omega \, \Theta(\alpha)
\ee
for all $\alpha \in \hat{\Omega}^1(\A)$ and $\omega \in \hat{\Omega}^r(\A)$.
Here $\pi$ denotes more generally the canonical projection
$\hat{\Omega}(\A) \oA \hat{\Omega}^1(\A) \rightarrow \hat{\Omega}(\A)$.
\vskip.1cm

If $(\hat{\Omega}(\A),\d)$ is a (generalized) differential calculus associated
with automorphisms $\{ \phi_s \, | \, s \in S \}$ of $\A$, we set
\be
    \V_s( \theta^{s'} )
  = \sum_{s'' \in S} \phi_s^{-1}( V^{s'}_{s,s''} ) \, \theta^{s''}
    \label{Vs_coeff}
\ee
with $V^{s'}_{s,s''} \in \A$. If the differential calculus is of the kind described
in section~\ref{sec:dc_auto}, using (\ref{vartheta_t}), (\ref{d_omega}), (\ref{na_E}) and
(\ref{VVs}), the torsion is given by
\be
      \Theta(\theta^s)
   = \theta^s \vartheta - \Delta(\theta^s)
      + \sum_{s' \in S} \theta^{s'} \, \V_{s'}(\theta^s)
   = \sum_{s',s'' \in S} ( \delta^s_{s'} - \delta^s_{s's''}
           + V^s_{s',s''} ) \, \theta^{s'} \theta^{s''}  \; .
\ee
Here we have chosen $t_s = 1$ for simplicity.
According to the decomposition of a 2-form into biangle, triangle and
quadrangle parts, the condition of vanishing torsion splits into
three sets of equations. The biangle torsion vanishes iff
\be
    V^s_{s',s''} = - \delta^s_{s'} \, \mathbf{1} \quad
    \forall s',s'' \in S \mbox{ with } s's''=e \; .
\ee
The vanishing of the triangle part of the torsion 2-form amounts to
\be
    V^s_{s',s''} = ( \delta^s_{s's''} - \delta^s_{s'} ) \, \mathbf{1} \quad
    \forall s',s'' \in S \mbox{ with } s's'' \in S \; .
\ee
In particular, this restricts the biangle and triangle components of
$V^s_{s',s''}$ to values in $\Bbbk$.
In reading off the quadrangle torsion components, one has to take
care of the 2-form relations (\ref{2formrel}). Since all these
relations are linear in $V^s_{s',s''}$, they do not feel the algebra
structure. In case of vanishing torsion, only some of the quadrangle
components of $V^s_{s',s''}$ could have values in $\A \setminus \Bbbk \mathbf{1}$.
\vskip.1cm

We should also mention that the condition of vanishing torsion has not
the distinguished status as in continuum differential (Riemannian)
geometry, as demonstrated in Ref.~\citen{DMH02b}.
\vskip.2cm

\begin{example}
In example~\ref{ex:qplane} we constructed differential calculi from
$\mathbb{Z}^2$ automorphism actions on the quantum plane algebra.
With respect to such a calculus, a linear connection on the
quantum plane is torsion-free iff
$V^{\hat{1}}_{\hat{2},\hat{1}} = V^{\hat{1}}_{\hat{1},\hat{2}}+1$
and $V^{\hat{2}}_{\hat{2},\hat{1}} = V^{\hat{2}}_{\hat{1},\hat{2}}-1$.
These are precisely the relations which we obtained for the group
lattice $(\mathbb{Z}^2, S = \{ \hat{1}, \hat{2} \})$ in
Ref.~\citen{DMH02b}, section 4.2.2. The only difference is that now
$V^s_{s',s''}$ takes values more generally in the quantum plane algebra.
\end{example}

\section{Noncommutative geometries associated with
         differentiable automorphisms}
\label{sec:ncg}
\setcounter{equation}{0}
In this section we assume that $\{ \phi_s, \phi_s^{-1} \, | \, s \in S \}$ is
a set of differentiable automorphisms of $\A$. Let $(\hat{\Omega}(\A),\d)$ be a
differential calculus which is inner at first order and satisfies
(\ref{theta_f_phi}) with respect to a basis $\theta^s$ of $\hat{\Omega}^1(\A)$.

\subsection{A semi-left-linear tensor product}
Let $T$ an arbitrary element of an $n$-fold tensor product space
$(\hat{\Omega}^1)^{\otimes n} := \hat{\Omega}^1(\A) \oA \ldots \oA \hat{\Omega}^1(\A)$.
Then a new tensor product can be defined via
\be
     \theta^s \oL T
  := \theta^s \oA \phi_s^{-1} T  \; .  \label{theta-oL}
\ee
This product is associative and has the following
semi-left-linearity property (with respect to the
1-form basis $\theta^s$):
\be
   f \, \theta^s \oL (f' \, T) = f f' \, \theta^s \oL T  \; .
\ee
If $\A$ is not commutative, this product is not left $\A$-linear
since $\alpha \oL (f \, T) \neq f \, \alpha \oL T$ for
$\alpha \in \hat{\Omega}^1$ and $f \in \A$, in general.

\subsection{Metric and compatibility with a linear connection}
Let $\nabla$ be a linear connection with transport operators $\V_s$.
The following procedure extends it to a connection on
$\hat{\Omega}^1(\A) \oL \hat{\Omega}^1(\A)$ (and more generally on
$n$-fold semi-left-linear tensor products). We define
\be
   \V_s(\alpha \oL \beta) := \V_s(\alpha) \oL \V_s(\beta)
\ee
for all $\alpha, \beta \in \hat{\Omega}^1(\A)$. The new $\V_s$ satisfy (\ref{VsfE})
and define via (\ref{VVs}) a left $\A$-module homomorphism
$\hat{\Omega}^1(\A) \oL \hat{\Omega}^1(\A) \rightarrow
 \hat{\Omega}^1(\A) \oA (\hat{\Omega}^1(\A) \oL \hat{\Omega}^1(\A))$
and thus, according to Lemma~\ref{lemma:connE}, a connection which
we also denote as $\nabla$.\footnote{Via (\ref{theta-oL})
a connection on $\hat{\Omega}^1(\A) \oL \hat{\Omega}^1(\A)$ defines a
connection on $\hat{\Omega}^1(\A) \oA \hat{\Omega}^1(\A)$ and vice versa.
In Refs.~\citen{lincon} a left $\A$-module map
$\sigma : \hat{\Omega}^1(\A) \oA \hat{\Omega}^1(\A) \rightarrow
 \hat{\Omega}^1(\A) \oA \hat{\Omega}^1(\A)$, which generalizes the
permutation map, has been used to extend a linear connection from
$\hat{\Omega}^1(\A)$ to $\hat{\Omega}^1(\A) \oA \hat{\Omega}^1(\A)$ by
$\nabla(\theta^s \oA \theta^{s'}) := \nabla \theta^s \oA \theta^{s'}
+ (\sigma \oA \mbox{id})(\theta^s \oA \nabla \theta^{s'})$. Though this is
very different from our approach, the problems addressed in Refs.~\citen{lincon}
can also be treated in our framework.}
\vskip.1cm

A \emph{metric} is taken to be an element
$\mathsf{g} \in \hat{\Omega}^1(\A) \oL \hat{\Omega}^1(\A)$, i.e.,
\be
    \mathsf{g} = \sum_{s,s' \in S} \g_{s,s'} \, \theta^s \oL \theta^{s'}
      \label{metric}
\ee
with a matrix of coefficients $\g_{s,s'} \in \A$ subject to suitable restrictions.
There is a distinguished class of metrics which are \emph{invariant} under the
action of all the automorphisms $\phi_s$, i.e.
\be
    \phi_s(\mathsf{g}) = \mathsf{g} \qquad \forall s \in S \; .
                         \label{invar_g}
\ee

A linear connection $\nabla$ is \emph{compatible} with the metric $\mathsf{g}$ if
\be
    \nabla \mathsf{g} = 0 \; .
\ee
This is equivalent to $\V_s(\mathsf{g}) = \mathsf{g}$ for all $s \in S$.
Using (\ref{Vs_coeff}), it becomes
\be
   \phi_s( \g_{s_1,s_2} ) = \sum_{s',s'' \in S} \g_{s',s''} \, V^{s'}_{s,s_1} \,
   V^{s''}_{s,s_2} \qquad \forall s,s_1,s_2 \in S \; .  \label{metric_compat}
\ee
In the special case where $\g_{s,s'} \in {\cal Z}(\A)$, this can be
written in matrix form as $\phi_s(\g) = V_s^T \g V_s$.
\vskip.1cm

A torsion-free linear connection which is compatible with a metric
$\mathsf{g}$ is called a \emph{Levi-Civita connection} of the respective
metric. The conditions of vanishing torsion and compatibility with a
given metric need not have any solution. If they do, it need not be unique
\cite{DMH02b}.
\vskip.2cm

\begin{example}
Let $(\hat{\Omega}(\A),\d)$ be the bicovariant differential calculus on
$GL_{p,q}(2)$ treated in section~\ref{sec:GLpq(2)}. A metric can then
be expressed in the form (\ref{metric}) with respect to the basis of
1-forms given in (\ref{GLpq(2)_thetas}).
Using (\ref{GLpq(2)_phithetas}), the invariance condition (\ref{invar_g})
imposes the following conditions on the metric components which we assume
to be symmetric:
$\phi_s(\g_{s',s''}) = \g_{s',s''}$ if $s',s'' = 1,3,4$,
$\phi_s(\g_{2,s'}) = r \, \g_{2,s'}$ if $s' = 1,3,4$, and
$\phi_s(\g_{2,2}) = r^2 \, \g_{2,2}$.
Here the action of $\phi_s$ on $\A$ is determined by (\ref{phi_alpha}).
In order to obtain a non-degenerate metric, some of its components must
lie in $\A \setminus {\cal Z}(\A)$.
(\ref{metric_compat}) determines corresponding compatible linear connections.
\end{example}

\section{Conclusions}
\label{sec:conclusions}
\setcounter{equation}{0}
In section~\ref{sec:dc_auto} we gave a recipe to construct
(distinguished) differential calculi over an associative algebra
$\A$ from subsets of automorphisms of $\A$.
A differential calculus obtained in this way is actually just the
differential calculus associated with the group lattice (Cayley digraph)
$(G,S)$ according to Ref.~\citen{DMH02a}. It is simply carried over to
$\A$ via an action of $G$ by automorphisms of $\A$ (provided some
technical requirements are fulfilled).
Familiar differential calculi, like those on the quantum plane,
turned out to be of this kind. We may conclude that they are
in fact basically group lattice differential calculi! The Wess-Zumino
calculus on the quantum plane \cite{qplane} in this way corresponds
to a $\mathbb{Z}^2$ lattice. Indeed, this relation can even be
established in a more explicit way, see appendix~\ref{sec:app}.
\vskip.1cm

The correspondence with group lattices only becomes visible in a
special coframe $\{ \theta^s \, | \, s \in S \}$. It should be noticed
that the $\theta^s$ are not, in general, ordinary 1-forms. Either one
has to extend (or rather restrict) the algebra $\A$ by appending
further generators and relations, or one has to enlarge the
bimodule of 1-forms.
\vskip.1cm

Geometric structures like linear connections and metrics
are also easily carried over from the differential geometry of
the associated group lattice. More precisely, this refers to
expressions in terms of products of the $\theta^s$ with coefficients
in the field $\Bbbk$. Clearly, the resulting geometry over $\A$
is richer since the coefficients lie more generally in $\A$.
\vskip.1cm

Whenever an automorphism group of a noncommutative algebra $\A$ is given
(see Refs.~\citen{Alev+Cham92,Gome+EL02} for further examples),
following the recipe of section~\ref{sec:dc_auto} distinguished differential
calculi can be constructed and thus a natural basis for differential
geometric structures on these algebras. Much of the formalism
developed in Refs.~\citen{DMH02a,DMH02b} for group lattices
then generalizes to these algebras.
\vskip.1cm

We should stress, however, that the construction of differential calculi
in section~\ref{sec:dc_auto} does not exhaust the possibilities admitting
the structure (\ref{theta_f_phi}), as demonstrated in particular in
section~\ref{sec:GLpq(2)} with an example of a bicovariant differential
calculus on $GL_{p,q}(2)$. This led us to a generalization of the construction
of section~\ref{sec:dc_auto} involving twisted inner derivations.
\vskip.1cm

Many examples of noncommutative field theories are constructed by deformation
quantization of classical field theories (see \cite{Doug+Nekr01}, for example).
Basically this amounts to replacing the classical product of functions
by a Moyal $\star$-product, which results in a noncommutative algebra $\A$.
On such algebras one can still establish an action
of a (discrete) group. Suppose we start with a lattice gauge theory. One should
then try to extend the action of the discrete translations (some $\mathbb{Z}^n$)
to the deformed algebra (see Ref.~\citen{AMNS00}, for example).
Our work provides a basis for corresponding geometric models.

\renewcommand{\theequation} {\Alph{section}.\arabic{equation}}

\begin{appendix}
\section{Realizations of some differential calculi}
\label{sec:app}
\setcounter{equation}{0}
Let $\A = \A_1 \otimes \A_2$ with two associative algebras $\A_1, \A_2$
over a field $\Bbbk$, and $(\Omega_1(\A_1),\d_1)$ a differential calculus
over $\A_1$. Defining $\d (\omega_1 \otimes f_2) := (\d \omega_1) \otimes f_2$
for all $\omega_1 \in \Omega_1$ and $f_2 \in \A_2$, we obtain a differential
calculus over $\A$. In the special case where $\A_1$ is commutative,
the above tensor product yields an algebra isomorphism $\A_2 \mapsto \A$.
In this case, the above construction of a differential calculus
carries a differential calculus over the commutative algebra $\A_1$
to a differential calculus over the (in general noncommutative)
algebra $\A$, respectively $\A_2$ due to the isomorphism.
\vskip.2cm

\begin{app_example}
Let $u,v$ freely generate a commutative algebra and let $U,V$ be generators
of a quantum plane, i.e. $UV = q VU$ where $q$ is a fixed parameter. Then
\be
     x = u \otimes U \, , \qquad y = v \otimes V    \label{qplane_newxy}
\ee
also generate a quantum plane with $xy = q yx$. A differential calculus
over the commutative algebra is defined by
\be
   u \, \d u = pq \,\d u \, u \, , \quad
   v \, \d u = pq \, \d u \, v \, , \quad
   u \, \d v = \d v \, u + (pq-1) \, \d u \, v \, , \quad
   v \, \d v = pq \, \d v \, v \, .
\ee
In fact, introducing $w$ with $v=uw$, this becomes
\be
  u \, \d u = pq \, \d u \, u \, , \quad
  w \, \d u = \d u \, w \, , \quad
  u \, \d w = \d w \, u \, , \quad
  w \, \d w = pq \, \d w \, w \, ,
\ee
which is the ``lattice differential calculus'' of Ref.~\citen{DMHS93}.
It is easily verified that the differential calculus induced
on the tensor product of the two algebras satisfies the
relations (\ref{qplane_diff}) in terms of the generators
(\ref{qplane_newxy}). In this way the differential calculus on the
quantum plane originates from the lattice calculus over the commutative
algebra generated by two variables, which may be realized as the
algebra of functions on $\mathbb{Z}^2$. In this realization a field
has the form $\psi(x,y) = \sum_{m,n} \psi_{m,n}(u,w) \, U^m V^n$
which can also be interpreted as a map from $\mathbb{Z}^2$ to the
quantum plane algebra (generated by $U,V$). Field theory and
differential geometry on the quantum plane are then essentially
field theory and differential geometry on the lattice $\mathbb{Z}^2$.
\end{app_example}

\begin{app_example}
Again, let $u,v$ freely generate a commutative algebra, but now we
assume that $U,V$ satisfy the relations of the $h$-deformed plane:
$[U,V] = h V^2$. This time we define the differential calculus over the
commutative algebra by the relations
\be
  [ v, \d v] = [ v, \d u] = [ u , \d v] = 0 \, , \quad
  [ u, \d u] = (h+h')(\d v \, u - \d u \, v) \, .
\ee
Introducing $w$ with $u = v w$, we obtain the simpler structure
\be
  [ v, \d v] = [ v, \d w] = [ w, \d v] = 0 \, , \quad
  [ w, \d w] = -(h+h') \, \d w \; .
\ee
The differential calculus introduced on the tensor product
now satisfies the relations (\ref{hplane_diff}) with $r=1$, where
\be
   x = v \otimes U + u \otimes V \, , \qquad y = v \otimes V \; .
\ee
In this way the differential calculus on the $h$-deformed plane
originates from the natural differential calculus over the algebra of
functions on $\mathbb{R} \times \mathbb{Z}$. The conclusions are
similar to those of the previous example.
\end{app_example}

It is interesting to see that prominent examples of differential
calculi over noncommutative algebras can indeed be realized as a tensor
product of a differential calculus over a commutative algebra with the
noncommutative algebra itself. A lot of further examples can be
constructed by using the differential calculi over commutative
algebras presented in Refs.~\cite{DMH02a,BDMH95} and arbitrary
associative algebras.
\end{appendix}


\begin{thebibliography}{**}
\bibitem{Mado99} Madore J 1999 {\em An Introduction to Noncommutative
 Differential Geometry and its Physical Applications} (Cambridge: Cambridge University Press)
\bibitem{Kaehler} Kunz E 1986 {\em K\"ahler Differentials} (Braunschweig: Vieweg); \\
 Berndt R 2002 K\"ahler differentials and some applications in arithmetical geometry
 {\em Hamburger Beitr\"age zur Mathematik}, Heft 136
\bibitem{Karoubi} Karoubi M 1999 {\em M\'ethodes quantiques en topologie alg\'ebrique}, preprint;
 2001 Quantum methods in algebraic topology {\em Contemp. Math.} {\bf 279} 177--193;
 2000 Braiding of differential forms and homotopy types,
 {\em C. R. Acad. Sci. Paris} {\bf 331}, S\'erie I, 757--762; \\
 Karoubi M and Alvarez M S 2001 Twisted K\"ahler differential forms, to appear
 in {\em J. Pure Appl. Algebra}
\bibitem{qcalc} Manin Yu I 1992 Notes on quantum groups and quantum
 de Rham complexes {\em Theor. Math. Phys.} {\bf 92} 997--1019; \\
 Dimakis A and M\"uller-Hoissen F 1992 Quantum mechanics on a lattice and q-deformations
 {\em Phys. Lett. B} {\bf 295} 242--248
\bibitem{DMH02a} Dimakis A and M\"uller-Hoissen F 2003 Differential geometry
 of group lattices {\em J. Math. Phys.} {\bf 44} 1781--1821 [math-ph/0207014]
\bibitem{DMH02b} Dimakis A and M\"uller-Hoissen F 2003
 Riemannian geometry of bicovariant group lattices
 {\em J. Math. Phys.} {\bf 44} 4220--4259 [math-ph/0212054]
\bibitem{Dima+Mado96} Dimakis A and Madore J 1996 Differential
 calculi and linear connections {\em J. Math. Phys.} {\bf 37} 4647--4661 [q-alg/9601023]
\bibitem{Woro89} Woronowicz S L 1989 Differential calculus on compact
 matrix pseudogroups (quantum groups) {\em Commun. Math. Phys.} {\bf 122} 125--170
\bibitem{Klim+Schm97} Klimyk A U and Schm\"udgen K 1997 {\em Quantum Groups and Their Representations}
 (Berlin: Springer)
\bibitem{MH92} M\"uller-Hoissen F 1992 Differential calculi on the quantum
 group $GL_{p,q}(2)$ {\em J. Phys. A} {\bf 25} 1703--1734
\bibitem{dc_qg} Carow-Watamura U, Schlieker M, Watamura S and Weich W 1991
 Bicovariant differential calculus on quantum groups $SU_q(N)$ and $SO_q(N)$
 {\em Commun. Math. Phys.} {\bf 142} 605--641; \\
 Jur\v{c}o B 1991 Differential calculus on quantized simple Lie groups
 {\em Lett. Math. Phys.} {\bf 22} 177--186; \\
 Schm\"udgen K and Sch\"uler A 1995 Classification of bicovariant differential calculi
 on quantum groups of type A, B, C and D {\em Commun. Math. Phys.} {\bf 167} 635--670; \\
 Heckenberger I and Schm\"udgen K 1998 Classification of bicovariant differential
 calculi on the quantum groups $SL_q(n+1)$ and $Sp_q(2n)$ {\em J. reine angew. Math.}
 {\bf 502} 141--162
\bibitem{Schm98} Schm\"udgen K 1998 On the construction of covariant differential
 calculi on quantum homogeneous spaces {\em J. Geom. Phys.} {\bf 30} 23--47; \\
 Hermisson U 1998 Construction of covariant differential calculi on quantum
 homogeneous spaces {\em Lett. Math. Phys.} {\bf 46} 313--322
\bibitem{DMHS93} Dimakis A, M\"uller-Hoissen F and Striker T 1993
 Non-commutative differential calculus and lattice gauge theory
 {\em J. Phys. A} {\bf 26} 1927--1949;
 1993 From continuum to lattice theory via deformation of the differential
 calculus {\em Phys. Lett. B} {\bf 300} 141--144
\bibitem{DMH94} Dimakis A and M\"uller-Hoissen F 1994 Discrete differential calculus:
 graphs, topologies, and gauge theory {\em J. Math. Phys.} {\bf 35} 6703--6735
\bibitem{Alev+Cham92} Alev J and Chamarie M 1992 D{\'e}rivations et automorphismes
 de quelques alg{\`e}bres quantiques {\em Comm. Algebra} {\bf 20} 1787--1802
\bibitem{Schi91} Schirrmacher A 1991 The multiparametric deformation of $GL(n)$
 and the covariant differential calculus on the quantum vector space
 {\em Z. Phys. C} {\bf 50} 321--328
\bibitem{qplane} Pusz W and Woronowicz S L 1989 Twisted second quantization
 {\em Rep. Math. Phys.} {\bf 27} 231--257; \\
 Wess J and Zumino B 1990 Covariant differential calculus on the quantum hyperplane
 {\em Nucl. Phys. (Proc. Suppl.) B} {\bf 18} 302--312; \\
 Brzezi{\'n}ski T, Dabrowski H and Rembieli{\'n}ski J 1992 On the
 quantum differential calculus and the quantum holomorphicity
 {\em J. Math. Phys.} {\bf 33} 19--24; \\
 Schm\"udgen K and Sch\"uler A 1993 Covariant differential calculi on quantum
  spaces and on quantum groups {\em C. R. Acad. Scie. Paris} {\bf 316} 1155--1160
\bibitem{Agha93} Aghamohammadi A 1993 The 2-parametric extension of $h$
 deformation of $GL(2)$, and the differential calculus on its quantum plane
 {\em Mod. Phys. Lett. A} {\bf 8} 2607--2613 [hep-th/9306071]
\bibitem{lincon} Mourad J 1995 Linear connections in non-commutative geometry
 {\em Class. Quantum Grav.} {\bf 21} 965--974; \\
 Dubois-Violette M, Madore J, Masson T and Mourad J 1995
 Linear connections on the quantum plane
 {\em Lett. Math. Phys.} {\bf 35} 351--358 [hep-th/9410199]; \\
 Georgelin Y, Masson T and Wallet J-C 1996
 Linear connections on the two parameter quantum plane {\em Rev. Math. Phys.} {\bf 8}
 1055--1060 [q-alg/9507032]; \\
 Aghamohammadi A, Khorrami M and Shariati A 1997
 $SL_{h}(2)$-symmetric torsionless connections {\em Lett. Math. Phys.}
 {\bf 40} 95--99; \\
 Cho S, Madore J and Park K S 1998
 Noncommutative geometry of the h-deformed quantum plane
 {\em J. Phys. A: Math. Gen.} {\bf 31} 2639--2654 [q-alg/9709007]
\bibitem{Gome+EL02} G{\'o}mez-Torrecillas J and Kaoutit L EL 2002
 The group of automorphisms of the coordinate ring of quantum symplectic space
 {\em Contrib. to Algebra and Geometry} {\bf 43} 597--601
\bibitem{Doug+Nekr01} Douglas M R and Nekrasov N A 2001 Noncommutative field theory
 {\em Rev. Mod. Phys.} {\bf 73} 977--1029 [hep-th/0106048]
\bibitem{AMNS00} Ambj{\o}rn J, Makeenko Y M, Nishimura J and Szabo R J 2000
 Lattice gauge fields and discrete noncommutative Yang-Mills theory
 {\em JHEP} {\bf 0005} 023 [hep-th/0004147]; \\
 Szabo R J 2001 Discrete noncommutative gauge theory 2001 {\em Mod. Phys. Lett. A}
 {\bf 16} 367--386 [hep-th/0101216]
\bibitem{BDMH95} Baehr H, Dimakis A and M\"uller-Hoissen F 1995
 Differential calculi on commutative algebras
 {\em J. Phys. A} {\bf 28} 3197--3222 [hep-th/9412069]
\end{thebibliography}
\end{document}